\documentclass{article}
\usepackage{amsmath,amsfonts,setspace,cancel,url,hyperref,verbatim,color,tikz,mathtools}
\usepackage{cite}
\usepackage[a4paper]{geometry}
\newcommand{\be}{\begin{equation}}
\newcommand{\ee}{\end{equation}}

\newcommand{\SL}[1]{\mathrm{SL}( #1 )}
\newcommand{\Sl}{\mathrm{SL}}

\newcommand{\SLf}{\mathrm{SL}(5)}
\newcommand{\so}[1]{\mathrm{SO}( #1 )}
\newcommand{\ODD}{\mathrm{O}(D,D)}

\newcommand{\gld}{{L}}
\newcommand{\hpartial}{\hat{\partial}}
\newcommand{\bpartial}{\bar{\partial}}
\newcommand{\bmu}{\bar{\mu}} 
\newcommand{\bnu}{\bar{\nu}} 
\newcommand{\brho}{\bar{\rho}} 
\newcommand{\blambda}{\bar{\lambda}} 
\newcommand{\bi}{\bar{i}} 
\newcommand{\bj}{\bar{j}} 
\newcommand{\bk}{\bar{k}} 
\newcommand{\cl}{\mathcal{L}}
\newcommand{\dm}{|m|^{-1}}

\newcommand{\tGamma}{\tilde{\Gamma}}

\newcommand{\tpartial}{\tilde{\partial}}
\newcommand{\nT}{\mathcal{T}}
\newcommand{\U}{U}
\newcommand{\Uxyw}{U_{xyw}} 
\newcommand{\Uxyz}{U_{xyz}} 
\newcommand{\Uyzw}{U_{yzw}} 
\newcommand{\Uxzw}{U_{xzw}} 
\numberwithin{equation}{section}
\begin{document}

\begin{titlepage}

\begin{center}

\hfill DAMTP-2014-88\\
\hfill  QGASLAB-14-05

\vskip 1.5cm

{\LARGE \sc Geometry and fluxes of $\mathrm{SL}(5)$ exceptional field theory}

\vskip 1cm

{\large \sc Chris D. A. Blair$^*$ and Emanuel Malek$^\dagger$} \\

\vskip 25pt

{\em  $^*$Department of Applied Mathematics and Theoretical Physics,\\Centre for Mathematical Sciences, University of Cambridge, \\
Wilberforce Road, Cambridge CB3 0WA, United Kingdom. \\ \vskip 5pt
$^\dagger$ Laboratory for Quantum Gravity and Strings, \\ Department of Mathematics and Applied Mathematics, University of Cape Town, \\ Private Bag X1, Rondebosch 7701, South Africa.\vskip 15pt }

{{\tt C.D.A.Blair@damtp.cam.ac.uk, \quad Emanuel.Malek@uct.ac.za\quad }} \\

\end{center}

\vskip 0.5cm

\begin{center} {\bf Abstract}\\[3ex]
\end{center}

\noindent 
We use a geometric approach to construct a flux formulation for the $\mathrm{SL}(5)$ U-duality manifest exceptional field theory. The resulting formalism is well-suited for studying gauged supergravities with geometric and non-geometric fluxes. Here we describe all such fluxes for both M-theory and IIB supergravity including the Ramond-Ramond fields for compactifications to seven dimensions. We define the locally non-geometric ``$R$-flux'' and globally non-geometric ``$Q$-flux'' for M-theory and find a new locally non-geometric $R$-flux for the IIB theory. We show how these non-geometric fluxes can be understood geometrically and give some examples of how they can be generated by acting with dualities on solutions with geometric or field-strength flux.

\thispagestyle{empty}

\end{titlepage}

\newpage
\tableofcontents 

\section{Introduction}

Flux compactifications of 10- and 11-dimensional supergravity are of huge phenomenological
importance. They provide a mechanism for moduli stabilisation, and one may also hope to use them to
realise deSitter and inflationary vacua \cite{Grana:2005jc}. They can also yield backgrounds with interesting holographic duals. 

There are numerous no-go theorems that make a simple flux compactification for stable deSitter and inflation impossible \cite{Hertzberg:2007wc,Caviezel:2008tf,Danielsson:2009ff}. A possible remedy for this situation without the need for elaborate brane set-ups may be provided by non-geometric backgrounds \cite{deCarlos:2009fq,deCarlos:2009qm,Blaback:2013ht,Damian:2013dq,Damian:2013dwa,Hassler:2014mla}.

In such a background the internal space of the compactification is patched by the T- or more generally U-duality symmetries of string theory. Although non-geometric backgrounds may look non-periodic and non-smooth from a spacetime perspective, they are well-defined backgrounds for the string, i.e. the string worldsheet on these backgrounds is a CFT. Furthermore, many non-geometric backgrounds can be obtained by duality transformations of geometric backgrounds \cite{Shelton:2005cf}. Beyond their potential phenomenological significance, non-geometric backgrounds are also interesting in their own right as they make explicit use of the stringy duality symmetries. This allows them to probe stringy regimes beyond supergravity.

Exceptional field theory exhibits the string dualities as manifest symmetries and hence is a natural language to describe non-geometric backgrounds. In this approach, extra coordinates are introduced which are thought of as conjugate to the wrapping modes of branes. U-duality then acts geometrically on the extended space given by the usual coordinates together with these new winding coordinates. Although we have extra coordinates, any physical field is constrained by the ``section condition'' to depend only on a subset of coordinates. 

Attempts to make dualities manifest in such a manner first appeared nearly 25 years ago \cite{Duff:1989tf, Duff:1990hn, Tseytlin:1990nb, Siegel:1993th, Siegel:1993xq}, and have intensified following the incorporation of the ideas of generalised geometry \cite{Gualtieri:2003dx,Hitchin:2004ut}, leading to a great deal of recent work realising T-duality and U-duality in a generalised or extended geometry \cite{
Hull:2009mi,
Hull:2007zu, Hillmann:2009ci,
Berman:2010is,
Berman:2013eva,
Aldazabal:2013sca
}.

Non-geometric fluxes have been studied extensively for the NS-NS sector of 10-dimensional supergravity \cite{Shelton:2005cf,Kachru:2002sk,Andriot:2011uh, Blumenhagen:2012nt, Andriot:2012wx, Andriot:2012an, Andriot:2013xca, Blumenhagen:2013aia, Blumenhagen:2013hva, Andriot:2014qla}. There one finds two non-geometric fluxes. Firstly, there is a globally non-geometric ``$Q$-flux'' which arises when the background can be locally described by some metric and antisymmetric Kalb-Ramond form which are however globally ill-defined. The metric and Kalb-Ramond form are globally well-defined, however, upon patching by a T-duality transformation. There is also a locally non-geometric ``R-flux'' which cannot be described, even locally, using a metric and Kalb-Ramond form. These backgrounds have a natural description through double field theory where T-duality is promoted to a manifest symmetry. Furthermore, non-geometric branes, also known as ``exotic branes'' \cite{deBoer:2010ud, deBoer:2012ma}, show signs of non-commutativity and even non-associativity \cite{Lust:2010iy,Blumenhagen:2010hj, Blumenhagen:2011ph,Condeescu:2012sp, Mylonas:2012pg,
Andriot:2012vb,Bakas:2013jwa,
Davidovic:2013rma,
Hassler:2013wsa,Blair:2014kla
}.
However, it is not clear how these results may generalise to M-theory, or are modified in the presence of R-R fields. 

The first aim of this paper is to provide a generalised geometric structure which naturally gives the Lagrangian of exceptional field theory. This will be based on a generalised torsion tensor of a flat connection from which one can uniquely produce the correct Lagrangian. This formalism turns out to be exceptionally useful for studying geometric and non-geometric fluxes.

The second aim of this paper is to use the ``flux formulation'' just constructed to describe the non-geometric fluxes of M-theory and IIB supergravity. We give definitions for globally and locally non-geometric fluxes for M-theory by identifying the spacetime tensors that appear in the embedding tensor of the lower-dimensional gauged supergravity. Intriguingly, the locally non-geometric ``$R$-flux'' in M-theory is not fully antisymmetric, in contrast to the NS-NS ``$R$-flux''. In the IIB theory, the Ramond-Ramond non-geometric sector has already been studied in \cite{Aldazabal:2010ef}. Here we extend that work in two ways. Firstly, we describe the locally non-geometric fluxes and find a new kind of ``$R$-flux'' which mixes the NS-NS and R-R sectors. Furthermore, we describe how the non-geometric fluxes can be understood geometrically, i.e. in terms of spacetime tensors, for both M-theory and IIB supergravity, generalising such work for the NS-NS sector as in \cite{Andriot:2011uh, Blumenhagen:2012nt, Andriot:2012wx, Andriot:2012an, Andriot:2013xca, Blumenhagen:2013aia,Andriot:2014qla}.

We will focus here on the exceptional field theory with manifest $\SLf$ duality, relevant for the scalar sector of compactifications to seven dimensions. This theory was originally introduced in \cite{Berman:2010is} and further studied in
\cite{Berman:2011cg,
Berman:2011jh,
Blair:2013gqa}. 
We note that for our purposes it is sufficient to focus solely on the scalar sector although it is also possible to treat the full 11-dimensional theory without making a truncation as for example in
\cite{Hohm:2014fxa
},
and to include fermions, as has been carried out for $E_7$ \cite{Godazgar:2014nqa}.

This restriction to the $\SLf$ theory allows us to explore fully the consequences of the extended
theory in a simpler setting than the higher U-duality groups (in particular, one does not yet need to
worry about dualisations of the M-theory three-form). We note that in string theory the
prototypical toy model of a situation leading to non-geometric flux is a three-torus carrying $H$-flux \cite{Kachru:2002sk}. 
The analogous M-theory situation would involve flux of the field strength of the three-form through a four-torus. This picks out $D=4$ as the lowest dimension in which one can study the M-theory versions of non-geometric fluxes: the duality group acting on four dimensions is of course $\SLf$. 

Although originally formulated for 11-dimensional supergravity, the $\SLf$ exceptional field theory also contains a reduction to (a truncation of) type IIB supergravity \cite{Blair:2013gqa} (see \cite{Hohm:2013pua} for a discussion for other duality groups). This is achieved by virtue of the fact that the fundamental field in the extended theory is the generalised metric: this can be parametrised in terms of physical fields in different ways. This will be an extremely important and useful fact for us when we want to find expressions for all possible non-geometric fluxes, for which one has to introduce alternative fields either instead of or alongside the usual parametrisations, similar to what has been done in string theory, for example in \cite{Grana:2008yw} and \cite{Andriot:2011uh}.

The geometry of extended field theories has been the subject of previous work 
\cite{ Aldazabal:2013mya, Cederwall:2013naa, Park:2013gaj} (and see also \cite{Coimbra:2011ky,Coimbra:2012af} for the case of exceptional generalised geometry, where the base manifold is not extended but the tangent bundle is). An interesting feature, reminiscent of issues in double field theory
\cite{
Hohm:2010xe,
Jeon:2010rw,
Hohm:2011si,
Berman:2013uda
},
is that there are obstructions to using the usual notions of Riemann and Ricci curvatures. Indeed, it has proven impossible to provide a definition for a generalised Riemann tensor for the exceptional extended geometry.
One can still define a generalised \emph{Ricci} tensor, leading to a Ricci scalar which can be used as the action. It is also possible to construct metric compatible connections which reproduce the known actions via the generalised curvature scalar: however, these connections seem unavoidably to contain undetermined components or else behave covariantly only under certain projections. 

This situation is entirely analogous to the doubled case. There, one proposed alternative \cite{Berman:2013uda} was to turn aside from attempting to build the action from generalised curvature, and instead to use a formalism in which a physically determined connection with non-vanishing generalised \emph{torsion}, and vanishing generalised curvature, led to the action. 

The outline of our paper is then as follows. After reviewing the $\SLf$ theory in section 2, we will show in section 3 how the torsionful geometric framework extends to the $\SLf$ exceptional field theory. Choosing as our covariant derivative the Weitzenb\"ock connection, we can uniquely fix the action in terms of the generalised torsion by demanding invariance under the local generalised Lorentz symmetry of the theory. 

We then study the geometrical content encoded by this connection, for the M-theory and IIB cases. The generalised torsion of the Weitzenb\"ock connection may be viewed as containing information about fluxes 
\cite{Geissbuhler:2011mx,Geissbuhler:2013uka 
}. By using the extended formalism, we are able to obtain all geometric and non-geometric fluxes: in order to do so we include dual fields and allow for the possibility of non-trivial derivatives in winding directions. The precise framework in which this should be possible in extended field theories is that of a Scherk-Schwarz compactification 
\cite{Scherk:1979zr, Hull:2006tp, Aldazabal:2011nj, Geissbuhler:2011mx , Grana:2012rr,  Aldazabal:2013mya, Berman:2012uy, Hohm:2014qga, Baron:2014bya}, leading to gauged supergravity. In fact, the generalised torsion of our formalism corresponds directly to the embedding tensor of gauged supergravity.  

We give this analysis and definition of all fluxes for M-theory in section 4. We then highlight some simple examples of duality chains involving geometric and non-geometric fluxes in section 5. This procedure is repeated in section 6, where we define the IIB fluxes, and section 7, where we present some example duality chains in type IIB theory. 

The reader solely interested in the definitions of the fluxes, and their unification into a U-duality tensor in the exceptional field theory, is invited to study sections 3.1 to 3.3 for the generalised geometrical definitions, and then sections 4 and 6 for the M-theory and IIB fluxes respectively. 

We note that the paper \cite{Aldazabal:2013mya} considers dynamical fluxes for the $E_7$ theory. However these are still packaged into a description in terms of a torsion-free connection with curvature and undetermined components, which drop out of the final action. For an interesting recent use of the Weitzenb\"ock connection in the context of generalised diffeomorphisms, see \cite{Rosabal:2014rga}.

\subsection*{Index conventions}

Indices in the $\mathbf{10}$ of $\SLf$ are referred to as ``big'' indices, and denoted using capital Roman letters, $A,B,C$. Flat indices (transforming under the generalised Lorentz group) here will be denoted with a bar over them, $\bar{A}, \bar{B},\bar{C}$. 

Indices in the $\mathbf{5}$ of $\SLf$ are referred to as ``little'' indices, and denoted using lower-case Roman letters from the start of the alphabet, $a,b,c$. The corresponding flat indices will be taken to be Greek, $\alpha,\beta,\gamma$.

Indices in the M-theory decomposition are four-dimensional spacetime indices, $i,j,k$, and four-dimensional flat indices, $\mu,\nu,\rho$.

In the IIB decomposition we have three-dimensional spacetime indices, $\mu,\nu,\rho$, as well as fundamental $\mathrm{SL}(2)$ indices, $i,j,k$. The corresponding flat indices will be denoted using bars.

\section{Review of the $\SLf$ theory}

We adopt here a top down approach to describing the $\SLf$ theory. From the 11-dimensional supergravity point of view, we describe solely what would be the scalar degrees of freedom appearing in a compactification to seven dimensions. This is a simplifying truncation which enables us to explore the essential consequences of the extended spacetime in a relatively clean set-up. 

\subsection{Generalised diffeomorphisms}

The $\SLf$ theory is defined on a 10-dimensional extended space \cite{Berman:2010is}. The coordinates $x^{A}$ lie in the antisymmetric 10-dimensional representation of $\SLf$ \cite{Berman:2011cg}. We write the 10-dimensional index $A$ as an antisymmetric pair of indices in the fundamental 5-dimensional representation of $\SLf$, $A \equiv [aa^\prime]$, $a,a^\prime=1,\dots,5$. 

The fundamental symmetry of the theory consists of generalised diffeomorphisms \cite{Berman:2011cg}. These are generated by a generalised vector $\U^{A}$ also in the $\mathbf{10}$ of $\SLf$. The general form of generalised diffeomorphisms is \cite{Berman:2012vc}
\be
\delta_\U V^A = \U^B \partial_B V^A - V^B \partial_B \U^A + Y^{AB}{}_{CD} V^C \partial_B \U^D \,,
\ee
where the $Y$-tensor is formed out of group invariants: in particular for $\SLf$ we have $Y^{AB}{}_{CD} = \epsilon^{e aa^\prime bb^\prime} \epsilon_{e cc^\prime dd^\prime}$, where $\epsilon_{abcde}$ is the totally antisymmetric invariant of $\SLf$. 

We can also give the explicit form of a generalised diffeomorphism acting on a fundamental $\SLf$ vector and covector as 
\be
\delta_\U V^a = 
\frac{1}{2} \U^{ef} \partial_{ef} V^a
+ \frac{1}{4} V^a \partial_{ef} \U^{ef}
- V^e \partial_{ef} \U^{af} \,,
\ee
\be
\delta_\U V_a  = 
\frac{1}{2} \U^{ef} \partial_{ef} V_a
- \frac{1}{4} V_a \partial_{ef} \U^{ef}
+ V_e \partial_{af} \U^{ef} \,.
\ee
This defines a generalised Lie derivative, $\delta_\U V^c \equiv \gld_\U V^c$, if we also take a scalar $\varphi$ to transform in the obvious manner, $\delta_\U \varphi = \frac{1}{2} \U^{ef} \partial_{ef} \varphi$.

The algebra of generalised Lie derivatives does not close unless one imposes the section condition \cite{Berman:2011cg}:
\be
\partial_{[ab } \otimes \partial_{cd]} = 0 \,,
\label{eq:sectioncondition}
\ee
where the pair of derivatives may act on any object or any pair of objects in the theory. Solving
the section condition amounts to choosing a lower-dimensional subspace of the 10-dimensional
extended space such that all quantities in the theory depend only on the coordinates of the
subspace, and so that \eqref{eq:sectioncondition} then holds. This choice of section thus amounts to picking out the ``physical'' space. 

The section condition is crucial in making statements about tensorial properties. For instance, consider the derivative of a scalar $\varphi$. In ordinary geometry, this is automatically a tensor. Here, however, one finds that
\be
\delta_\U \partial_{ab} \varphi = \gld_\U \partial_{ab} \varphi  + 3 \partial_{[ab} \varphi \partial_{ef] } \U^{ef} \,.
\ee
The final terms vanish by the section condition. 

\subsection{The action} 

The bosonic fields of the theory live in a coset $\mathbb{R}^+ \!\times \SLf / \mathrm{SO}(5)$\footnote{The extra $\mathbb{R}^+$ factor is a consequence of our truncation, and leads to an extra scalar degree of freedom related to the warping of the ignored external seven directions, see for example \cite{Coimbra:2011ky,Blair:2013gqa}}., and in principle depend on the full ten-dimensional extended coordinates $x^{ab}$. They may be packaged into a ``generalised metric'' $M_{AB}$ \cite{Duff:1990hn,Hull:2007zu} which parametrises the given coset and serves as the metric on the extended spacetime \cite{Berman:2010is}. As a consequence of the coset condition this generalised metric $M_{AB}$ can be decomposed in terms of a ``little metric'' $m_{ab}$ \cite{Berman:2011cg}, with
\be
M_{AB} \equiv M_{aa^\prime, bb^\prime} = m_{ab} m_{a^\prime b^\prime} - m_{ab^\prime} m_{a^\prime b} \,,
\ee
where $m_{ab}$ is symmetric, and is a rank two tensor under $\SLf$ U-dualities. 

Note that although we will refer to $m_{ab}$ as the little metric it itself is not a metric on some space. However, it provides the most convenient way of constructing the theory, containing exactly the right number of degrees of freedom to parametrise the coset $\mathbb{R}^+ \!\times \SLf / \mathrm{SO}(5)$. We should also mention that one can only decompose the full generalised metric in this way in the $\SLf$ theory, and not for the higher exceptional groups.

The action for the truncated theory is completely fixed by searching for an expression quadratic in derivatives of the little metric which is a scalar under generalised diffeomorphisms up to section condition. It is given by \cite{Berman:2010is,Park:2013gaj}
\begin{equation}
 \begin{split}
  S &= \int_\Sigma |m|^{-1} \left( - \frac{1}{8} m^{ab} m^{a^\prime b^\prime}
\partial_{a a^\prime} m^{c d} \partial_{b b^\prime} m_{c d} + \frac{1}{2} m^{ab}
m^{a^\prime b^\prime} \partial_{aa^\prime} m^{cd} \partial_{cb^\prime} m_{bd}
\right.\\
& \quad \left. +\frac{1}{2} \partial_{aa^\prime} m^{ab} \partial_{bb^\prime}
m^{a^\prime
b^\prime} + \frac{3}{8}  m^{ab}  m^{a^\prime b^\prime}
\partial_{aa^\prime}\ln |m| \, \partial_{bb^\prime} \ln |m| - 2 m^{a^\prime
b^\prime} \partial_{aa^\prime} m^{ab} \partial_{bb^\prime} \ln|m| \right.
\\
& \quad \left. + m^{a^\prime
b^\prime}\partial_{aa^\prime}\partial_{bb^\prime} m^{ab} - m^{ab}
m^{a^\prime b^\prime} \partial_{a a^\prime} \partial_{bb^\prime} \ln |m| \right)
\,, \label{eqSL5action}
 \end{split}
\end{equation}
where $\Sigma$ is some lower-dimensional section of the full ten-dimensional
theory, and we have used the determinant of the little metric, $m \equiv \det m_{ab}$, to define an $\SLf$ singlet integral measure,
$|m|^{-1}$.
\subsection{Section choices}

Let us briefly discuss the two inequivalent sections, corresponding to (truncations of) 11-dimensional and 10-dimensional type IIB supergravity. We shall give explicit expressions for the decomposition later, when we evaluate the generalised fluxes. 

\vspace{1em}
\noindent\textbf{M-theory section}

The conventional solution to the section condition is the M-theory section \cite{Berman:2010is},
where we split the 5-dimensional index $a=i,5$ where $i$ becomes a 4-dimensional spacetime index.
One then takes all fields to depend only on the four coordinates $x^i \equiv x^{i5}$, and to have no
dependence on the $x^{ij}$. After choosing an appropriate parametrisation of the generalised metric
in terms of a metric $g_{ij}$, three-form gauge field $C_{ijk}$ and additional scalar $\phi$ one
find that the action \eqref{eqSL5action} reduces to a truncation of 11-dimensional supergravity to
four dimensions \cite{Berman:2010is,Park:2013gaj}. (This truncated theory essentially corresponds to the
internal (scalar) sector of 11-dimensional SUGRA reduced to seven dimensions, note however that in
this truncation we
keep the 4-dimensional coordinate dependence. Similar remarks apply in the IIB case below.)

A type IIA section may be trivially obtained from this choice by supposing that we are also independent of one of the four coordinates $x^i$, in the usual way. 

\vspace{1em}
\noindent\textbf{IIB section} 

An alternative section \cite{Blair:2013gqa} is given by making a $3+2$ split of the 5-dimensional index, $a=\mu,i$ where now $\mu$ becomes a 3-dimensional spacetime index and $i=1,2$ becomes a fundamental $\mathrm{SL}(2)$ index corresponding to the S-duality symmetry of type IIB.\footnote{Similar inequivalent IIB sections were also discussed in \cite{Hohm:2013pua} for the groups $E_6$, $E_7$ and $E_8$.} Our fields are taken to only depend on the three coordinates $x^{\mu \nu}$, and are independent of the other coordinates $x^{\mu i}$, $x^{ij}$. The spacetime coordinates $x^{\mu \nu}$ may be dualised to carry a single lower index, $\tilde x_\mu \equiv \frac{1}{2} \eta_{\mu \nu \rho} x^{\nu \rho}$, so that vectors in this parametrisation are written with lower indices. One may then parametrise the generalised metric by introducing a metric $g^{\mu \nu}$, a pair of two-forms $C^{\mu \nu i}$, a unit determinant two-by-two matrix of scalars $\mathcal{M}_{ij}$, which incorporates the Ramond-Ramond zero form and string dilaton, and again an additional scalar $\phi$. Evaluating the action in this section and parametrisation, one obtains a truncation of type IIB supergravity to three dimensions\cite{Blair:2013gqa}.  

\vspace{1em}
Although the parametrisations we have described here for the IIB and M-theory cases give the usual field content and description of these theories, other choices, involving so-called dual fields, are possible. These will be important later on. 

\section{Connections, torsion and the action}

In this section, we shall introduce geometric structure on the $\SLf$ theory, in the form of
connections. The goal is to seek some geometric origin of the action \eqref{eqSL5action}. This
problem has been considered before by other authors, both in the context of $\SLf$ and for other
duality groups
\cite{Park:2013gaj,Cederwall:2013naa,Aldazabal:2013mya}. We wish to provide an alternative approach,
which evades some of the issues that arise when considering metric-compatible connections with
curvature, and which is suited for describing fluxes. 

\subsection{Connections}

We introduce a covariant derivative in the $\SLf$ theory in the usual way, by seeking a connection
$\Gamma_{BC}{}^A$ which, given the form of generalised diffeomorphisms, must transform as
\be
\delta_\U \Gamma_{BC}{}^A = \gld_\U \Gamma_{BC}{}^A + \partial_B \partial_C \U^A - Y^{AD}{}_{CE} \partial_D \partial_B \U^E \,.
\ee
Here we have introduced a connection carrying solely ``big'' indices. For practical applications, it
is convenient to introduce instead a connection which acts not on the antisymmetric representation
but on the fundamental. This ``little'' connection is defined via
\footnote
{
For the sake of completeness, note that if a generalised vector $V^a$ also carries weight $\omega$, so that
\[
\delta_\U V^a = \gld_\U V^a + \frac{1}{2} \omega \partial_{ef} \U^{ef} V^a \,
\]
then its covariant derivative should be defined as 
\[
\nabla_{ab} V^c = \partial_{ab} V^c + \Gamma_{abd}{}^c V^d 
+ \omega ( \Gamma_{eab}{}^e - \Gamma_{eba}{}^e ) V^c \,,
\]
from which follows that if $V^{ab}$ has weight one that $\nabla_{ab} V^{ab} = \partial_{ab} V^{ab}$.
}
\be
\nabla_{ab} V^c \equiv \partial_{ab} V^c + \Gamma_{ab d}{}^c V^d \,,
\ee
and its transformation must be
\be
\delta_\U \Gamma_{abd}{}^c = \gld_\U \Gamma_{abd}{}^c - \delta^c_d \frac{1}{4} \partial_{ab} \partial_{ef} \U^{ef} + \partial_{ab} \partial_{de} \U^{ce} \,,
\label{eq:slflittletrans}
\ee
up to terms that vanish by the section condition. 

Given such a little connection there is an associated big connection, defined by
\be
\Gamma_{BC}{}^A \equiv \Gamma_{bb^\prime cc^\prime}{}^{aa^\prime} = 4 \Gamma_{bb^\prime [c}{}^{[a} \delta_{c^\prime]}^{a^\prime]} \,.
\label{eq:BigLittleConnection}
\ee
Now, in ordinary general relativity one can easily find a special connection which leads naturally
to the action. This is the Levi-Civita connection, which is the unique torsion-free
metric-compatible connection. In extended theories in general, matters are not quite so simple. 

Ideally, we would like to find a connection which
\begin{itemize}
\item provides a true covariant derivative, mapping tensors to tensors,
\item annihilates the generalised metric, $\nabla_{ab} m_{cd} = 0$,
\item also annihilates the $\SLf$ invariant $\epsilon_{abcde}$,
\item is completely determined in terms of the physical fields,
\item by analogy with general relativity, has vanishing generalised torsion (to be defined in the
next subsection),
\item has a natural curvature scalar that leads to the action \eqref{eqSL5action}. 
\end{itemize}
Unfortunately, it proves difficult to meet all these requirements. One issue that 
arises is simply how to generalise curvature. 
The normal definition of the Riemann tensor does not provide a generalised tensor. Interestingly, despite several attempts, it has proven impossible to construct a definition for a generalised Riemann tensor which is a true generalised tensor \cite{Park:2013gaj,Aldazabal:2013mya, Cederwall:2013naa}. One can still produce a two index tensor which is a generalised Ricci tensor: contracting this tensor with the generalised metric yields a generalised Ricci scalar which can be used as a Lagrangian.

However, when one now looks for explicit connections one is forced to sacrifice one of the above
requirements. This is very similar to the case of double field theory.
One can find a covariant derivative which transforms correctly, but which contains
undetermined components, not expressible in terms of the physical fields, as in  
\cite{Aldazabal:2013mya, Cederwall:2013naa}. Alternatively, one can produce a derivative which has
no undetermined components, but which only transforms covariantly in certain circumstances (and is
said to be semi-covariant) 
\cite{Park:2013gaj}. 

We stress that these apparent issues do not in fact cause any difficulty in obtaining the correct
action. One finds that the undetermined components, or equivalently those that do not transform
covariantly, in fact drop out when one constructs the generalised Ricci scalar in these approaches. 

Our goal in this section of the paper is to present an alternative framework, in which one does not
use notions of curvature but instead considers a torsionful flat connection. This is provided 
by the Weitzenb\"ock connection. 

\subsection{The generalised torsion}

First, let us show what we mean by generalised torsion. The generalised torsion of a connection is defined by replacing partial derivatives with covariant derivatives in the generalised Lie derivative:
\be
\gld_U (\nabla) V^A - \gld_U (\partial) V^A = \tau_{B C}{}^A U^B V^C \,, 
\ee
giving
\be
\tau_{BC}{}^A = \Gamma_{BC}{}^A - \Gamma_{CB}{}^A + Y^{AD}{}_{CE} \Gamma_{DB}{}^E \,.
\ee
Alternatively we may defined a generalised torsion in terms of fundamental quantities and a little connection:
\be
\gld_U (\nabla) V^a - \gld_U (\partial) V^a = \frac{1}{2} \tau_{bcd}{}^a U^{bc} V^d \,,
\ee
giving
\be
\tau_{bc d}{}^{a} = 3 \Gamma_{[bcd]}{}^a - \Gamma_{e[bc]}{}^e \delta^a_d - 2 \Gamma_{e d [ b}{}^e \delta_{c]}^a\,.
\label{eq:torsion}
\ee
For big and little connections related by \eqref{eq:BigLittleConnection}, the resulting big and little torsions are related in the same way
\be
\tau_{bb^\prime cc^\prime}{}^{aa^\prime} = 4 \tau_{bb^\prime [c}{}^{[a} \delta_{c^\prime]}^{a^\prime]} \,.
\ee
We may therefore choose to use either as the basis for our construction. It is more convenient to work with the little torsion.

Before proceeding, it will be useful to classify the transformation properties of the torsion under
global $\SLf$. A tensor $\tau_{bcd}{}^a$ with $\tau_{bcd}{}^a = - \tau_{cbd}{}^a$ lives in the
tensor product representation $\mathbf{5} \otimes \mathbf{\overline{5}} \otimes \mathbf{{10}} =
\mathbf{{10}} \oplus \mathbf{{10}} \oplus \mathbf{{15}} \oplus \mathbf{{40}} \oplus \mathbf{{175}}$.
The explicit realisation of the tensor decomposition into irreducibles is:
\be
\begin{split}
\tau_{bcd}{}^{a} &= \widetilde{T}_{[bcd]}{}^a
+ \frac{2}{3} \left( \widetilde{T}_{b(cd)}{}^a - \widetilde{T}_{c(bd)}{}^a \right) 
+ \frac{1}{9} \delta^a_d A_{bc} + \frac{5}{9} \delta^a_{[b} A_{c] d}
+  \frac{1}{2} \delta^a_{[b} S_{c]d} \\
& \quad + \frac{1}{3} \delta^a_d \tau_{bce}{}^e + \frac{2}{3} \delta^a_{[b} \tau_{c]de}{}^e \,,
\end{split}
\ee
where
\be
\widetilde{T}_{bcd}{}^a = T_{bcd}{}^a -  \frac{1}{2} \delta^a_{[b} S_{c]d}
- \frac{1}{9} \delta^a_d A_{bc} - \frac{5}{9} \delta^a_{[b} A_{c] d} 
\quad , \quad
 \widetilde{T}_{abc}{}^a = 0 = \widetilde{T}_{bca}{}^a\,,
\ee
with
\be
T_{bcd}{}^a = \tau_{bcd}{}^a
- \frac{1}{3} \delta^a_d \tau_{bce}{}^e - \frac{2}{3} \delta^a_{[b} \tau_{c]de}{}^e \quad , \quad
T_{bca}{}^a = 0\,,
\ee
and
\be
S_{cd} = T_{e(cd)}{}^e 
\quad , \quad
 A_{cd} = T_{e[cd]}{}^e \,.
\ee
The trace $\tau_{bce}{}^e$ lives in the $\mathbf{{10}}$, $\widetilde{T}_{[bcd]}{}^a$ in the $\mathbf{{40}}$, the symmetric $S_{cd}$ in the $\mathbf{{15}}$, the antisymmetric $A_{cd}$ in the other $\mathbf{{10}}$ and the mixed symmetry $\widetilde{T}_{b(cd)}{}^a$ in the $\mathbf{{175}}$.

For the torsion \eqref{eq:torsion} one finds that $A_{cd} = 0$ and
$\widetilde{T}_{b(cd)}{}^a = 0$, as well as
\be
S_{ab} = 2 \left( \Gamma_{eab}{}^e+ \Gamma_{eba}{}^e \right) \,,
\ee
\be
\tau_{abe}{}^e = \Gamma_{abe}{}^e - \frac{1}{2} \Gamma_{eab}{}^e + \frac{1}{2} \Gamma_{eba}{}^e \,,
\ee
\be
\widetilde{T}_{bcd}{}^a = \widetilde{T}_{[bcd]}{}^a = 3 \Gamma_{[bcd]}{}^a - \delta^a_{[b} \Gamma_{cd]e}{}^e - 2 \delta^a_{[b} \Gamma_{|e|cd]}{}^e \,.
\ee
Hence it contains just the irreducibles $\mathbf{10}$, $\mathbf{15}$ and $\mathbf{40}$. Note that
the latter two irreps are those of the embedding tensor of gauged maximal supergravity in
$7$-dimensions (where the duality group is of course $\mathrm{SL}(5)$) \cite{Samtleben:2005bp}. The
remaining $\mathbf{10}$ can be thought of as a trombone gauging. For convenience we relabel it as
$\tau_{ab} \equiv \tau_{abe}{}^e$.

\subsection{The Weitzenb\"ock connection}

First, we introduce a generalised vielbein for the little metric. Recall that this object
parametrised the coset $\mathbb{R}^+ \times \SLf/\mathrm{SO}(5)$\footnote{If we were dealing with a
truncation including the time direction, the coset space would instead be $\mathbb{R}^+ \times \SLf/
\mathrm{SO}(3,2)$ \cite{Hull:1998br}. However, in this paper we assume our truncation is
Euclidean.}. The group $\mathrm{SO}(5)$ acts by local internal rotations, and may be thought of as the generalised Lorentz group of the extended theory. We define a flat metric $m_{\alpha \beta}$, which we can take to be the identity, and introduce a generalised vielbein $E^\alpha{}_a$ such that
\be
m_{ab} = E^\alpha{}_a E^\beta{}_b m_{\alpha \beta} \,.
\ee
The flat index $\alpha$ then transforms under local $\mathrm{SO}(5)$ transformations. Note that we will use $m_{\alpha \beta}$ to lower and raise flat indices. 

This introduction of a ``little'' generalised vielbein is compatible with the existence of the big generalised metric. If we denote the big flat indices with bars, then the associated big generalised vielbein would be given by
\be
E^{\bar{A}}{}_A \equiv E^{\alpha \alpha^\prime}{}_{aa^\prime} = E^\alpha{}_a
E^{\alpha^\prime}{}_{a^\prime} - E^{\alpha^\prime}{}_a E^\alpha{}_{a^\prime} \,,
\ee
with the flat big generalised metric given by the expected formula, $M_{\bar{A} \bar{B}} \equiv m_{\alpha \beta} m_{\alpha^\prime \beta^\prime} - m_{\alpha \beta^\prime} m_{\beta \alpha^\prime}$. 

We may then define the generalised Weitzenb\"ock connection with little indices:
\be
\Omega_{bcd}{}^a =  E_\alpha{}^a \partial_{bc} E^\alpha{}_d \,.
\ee
This connection can be checked to annihilate both the little metric and the $\SLf$ invariant
$\epsilon_{abcde}$. It transforms as in \eqref{eq:slflittletrans} up to section condition. It has
non-vanishing generalised torsion, but has vanishing curvature and generalised curvature. This is
easiest to check by using the ``big'' form of the connection. Then, as in \cite{Berman:2013uda}, one
finds that although the ordinary expression for the Riemann tensor is not in general
a generalised tensor, it is for the Weitzenb\"ock connection by the section condition, and also
vanishes for this connection. Similarly, one can check that the proposed general form for a
generalised Ricci tensor \cite{Cederwall:2013naa} (see also \cite{Park:2013gaj, Aldazabal:2013mya})
then vanishes for the Weitzenb\"ock connection, again using the section condition. 

The associated big Weitzenb\"ock connection
\be
\Omega_{BC}{}^A = E_{\bar{A}}{}^A \partial_B E^{\bar{A}}{}_C \,,
\ee
is related to the little one by \eqref{eq:BigLittleConnection}. 

In order to use the generalised Weitzenb\"ock connection we need parallelisability in the sense of generalised geometry. Let us just mention that while parallelisability is a notoriously stringent requirement for manifolds, it is a more relaxed requirement here. This is because the generalised vielbein contains the spacetime vielbein as well as $p$-forms and even at points where the spacetime vielbein vanishes, the $p$-forms may be non-zero. Indeed, this allows spheres of all dimensions to be parallelisable \cite{Lee:2014mla}.\footnote{After this work first appeared, we were made aware of a proof of generalised parallelisability of hyperboloidal spaces to appear in the revised version of \cite{Baron:2014bya}.} The examples we consider later will be parallelisable in the generalised sense.

\subsection{Constructing the action}

The Weitzenb\"ock connection is not invariant under local generalised Lorentz transformations,
\be
E^\alpha{}_a \rightarrow \lambda^\alpha{}_\beta E^\beta{}_a \quad , \quad \lambda_{\alpha \beta} = - \lambda_{\beta\alpha} \,.
\ee
In order to construct a Lagrangian in terms of the generalised torsion of this connection, we can use this lack of invariance as a constraining principle. We are looking to write down all possible torsion squared terms but as the generalised torsion does not fall into an irreducible representation of $\SLf$, there are naively many possible such terms that can be written down. However, several of these are equivalent. This is made clearer by working in terms of the torsion irreducibles, in terms of which there are merely five independent terms quadratic in the torsion:
\be
\begin{split} 
     m^{ab} m^{cd} S_{ac} S_{bd} \,,\,
     m^{ab} m^{cd} S_{ab} S_{cd} \,,\,
     m^{ab} m^{cd} \tau_{ac} \tau_{bd} \,,\,
\\ 
     m_{ab} m^{cd} m^{ef} m^{gh} \tilde{T}_{ceg}{}^a \tilde{T}_{dfh}{}^b\,,\,
     m^{ab} m^{cd} \tilde{T}_{acf}{}^e \tilde{T}_{bde}{}^f\,,
\end{split} 
\ee
and a single term involving the covariant derivative of the torsion trace:\footnote{Note that for the Weitzenb\"ock connection, one has the useful result that
\[
\nabla_{ab} \dm = - 2 \dm \tau_{ab} \,.
\]
}
\be
m^{ab} m^{a'b'} \nabla_{aa'} \tau_{bb'} \,.
\ee
It is straightforward to vary each of these terms under generalised Lorentz transformations, with the result being that there is a unique (up to scale) combination of torsion squared contractions giving a generalised Lorentz invariant scalar up to section condition. This scalar is 
\be
\begin{split}
& \frac{1}{16} m^{ab} m^{cd} S_{ac} S_{bd}  - \frac{1}{32} (m^{ab}S_{ab})^2 + \frac{5}{3}
m^{ab}m^{cd} \tau_{ac} \tau_{bd}
\\ &+ \frac{1}{12}m_{ab}m^{cd} m^{ef} m^{gh} \widetilde{T}_{ceg}{}^a \widetilde{T}_{dfh}{}^b
+ \frac{1}{4} m^{ab} m^{cd} \widetilde{T}_{acf}{}^e \widetilde{T}_{bde}{}^f
 - 2 m^{ab}m^{cd} \nabla_{ac} \tau_{bd}\,,
\end{split}
\label{eq:SLft2Lag}
\ee
and under generalised Lorentz variation this has an anomalous transformation 
\be
\begin{split}
 6 m^{ab} E_\alpha{}^c E_\beta{}^d \Omega_{[ae|b|}{}^e \partial_{cd]} \lambda^{\alpha \beta} & =
- m^{ab} E_\alpha{}^c E_\beta{}^d \Big( \Omega_{eab}{}^e \partial_{cd} \lambda^{\alpha \beta}
+ 2 \Omega_{edb}{}^e \partial_{ac} \lambda^{\alpha \beta}
\\
& + 2 \Omega_{adb}{}^e \partial_{ce} \lambda^{\alpha \beta}
- \Omega_{cdb}{}^e \partial_{ae} \lambda^{\alpha \beta} \Big) \,,
\end{split}
\ee
which indeed vanishes by the section condition. 

It is then possible to check that this Lagrangian \eqref{eq:SLft2Lag} agrees with that appearing in the action \eqref{eqSL5action} up to the section condition term
\be
\begin{split}
& + \frac{1}{2}(  m^{a b} m^{c d} \Omega_{a e c}{}^{e} \Omega_{d f b}{}^f + 2 m^{a b} m^{c d} \Omega_{a c b}{}^e \Omega_{e f d}{}^f \\
 & - m^{a b} m^{c d} \Omega_{a e b}{}^e \Omega_{c f d}{}^f - m^{a b} m^{c d} \Omega_{a f c}{}^e \Omega_{d e b}{}^f + m^{a b} m^{c d} \Omega_{a f b}{}^e \Omega_{c e d}{}^f) \,.
\end{split}
\ee
This term can be written as
\be
\frac{1}{2} Y^{AB}{}_{CD} \Omega_{AE}{}^C \Omega_{BF}{}^D M^{EF} = - \frac{1}{8} \epsilon^{eaa^\prime b^\prime} \epsilon_{ecc^\prime dd^\prime} 
\Omega_{aa^\prime e}{}^c \Omega_{bb^\prime f}{}^d m^{c^\prime f} m^{d^\prime e} \,,
\ee
and can be seen to be identical to the term which in \cite{Berman:2012uy} was necessary to add in by hand in order to obtain a consistent Scherk-Schwarz reduction. This is exactly as expected from the double field theory case, where the Lagrangian resulting from requiring invariance under generalised Lorentz transformations led exactly to the analogous term needed for gauged double field theory \cite{Berman:2013uda}.

\subsection{Relationship to gauged supergravity} 

Let us briefly expand on the links to gauged supergravity mentioned above. Recall that in a gauged supergravity, some subgroup of the global duality group, which here is our $\mathrm{SL}(5)$, is enhanced to a local gauge symmetry. 

It is possible to formulate gauged supergravities in any dimension as deformations of the more familiar ungauged supergravities. Here, the embedding tensor \cite{Samtleben:2008pe,Nicolai:2000sc}, which describes explicitly the embedding of the gauged subgroup into the larger duality group, plays an important role. In order to preserve supersymmetry, this object obeys various constraints. Some of the allowed components of the embedding tensor correspond to gaugings that can be obtained via a Scherk-Schwarz or flux compactification. However, others do not - there are many allowed gaugings which can give a gauged supergravity which appear to have no higher dimensional interpretation. 

A resolution is provided by extended field theory. It turns out that Scherk-Schwarz compactifications of double field theory give, after solving a set of Scherk-Schwarz constraints that replace the section condition, the Lagrangians of gauged (half-maximal) supergravity \cite{Aldazabal:2011nj,Geissbuhler:2011mx,Grana:2012rr} (for reviews of this material see \cite{Aldazabal:2013sca, Berman:2013eva}). However, one can obtain all possible gaugings in this way. This relies crucially on the existence of the extra coordinates, which enter into the theory in form of generalised gaugings. Thus one finds that double field theory provides a higher dimensional uplift for all (electrically) gauged half-maximal supergravities. This has been extended to the gauged maximal supergravities in the case of the extended field theories for U-duality \cite{Berman:2012uy, Aldazabal:2013mya, Hohm:2014qga,Baron:2014bya}.

In a gauged Scherk-Schwarz reduction of the $\SLf$ exceptional field theory, one introduces twisting matrices $W_a^A$ which carry all dependence on the internal coordinates of the compactification. Here this would be the 10 coordinates $x^{ab}$ - the resulting effective theory will depend only on the external coordinates, which we denote $\mathbb{X}$. Quantities which depend only on $\mathbb{X}$ will be denoted with a hat, and the Scherk-Schwarz Ansatz is to assume that all physical fields and gauge parameters may be factorised as
\be
V^a ( x, \mathbb{X} ) = (W^{-1})_A{}^a (x ) \hat{V}^a ( \mathbb{X} ) \,.
\ee
Under this assumption, one finds that the symmetries of the theory are governed by the resultant decomposition of the generalised Lie derivative
\be
\gld_\U V^a = (W^{-1})_A{}^a \left( \hat{\gld}_{\hat{\U}} \hat{V}^A - \frac{1}{2} \tau_{BCD}{}^A \hat{\U}^{BC} \hat{V}^D \right) \,,
\ee
where $\hat{\gld}_{\hat{\U}}$ is just the generalised Lie derivative written in terms of only hatted quantities and using only the capital indices $A,B,C$, which are the indices of the gauged exceptional field theory. The quantity $\tau_{BCD}{}^A$ is then nothing but the generalised torsion \eqref{eq:torsion} written in terms of the quantities 
\be
\Omega_{BCD}{}^A \equiv ( W^{-1} )_D{}^e \partial_{BC} W^A{}_e \,.
\ee
The piece $- \frac{1}{2} \tau_{BCD}{}^A \hat{\U}^{BC} \hat{V}^D$ appearing in the local symmetries of the gauged theories then amounts to a gauging. We see a direct link here between the generalised torsion and the embedding tensor. 

There are various conditions that must still be imposed to ensure we have a consistent theory.
Firstly, as we want to interpret $\tau_{BCD}{}^A$ as giving effectively the structure constants for
some gauge group, we must assume that it is constant. One also has consistency conditions from
requiring these be preserved under the local symmetries, and from requiring closure of the algebra
of symmetries of the gauged theory. This gives various quadratic constraints on the torsion
\cite{Berman:2012uy}, which are one and the same as the quadratic constraints on the embedding
tensor of gauged supergravity \cite{Samtleben:2005bp}. 

Note that in our formulation, the only bosonic field is the little metric, which is decomposed in terms of the twists as
\be
m_{ab}(x,\mathbb{X}) = W^A{}_a (x) W^B{}_b (x)  \hat{m}_{AB} (\mathbb{X} ) \,.
\ee
The dynamical degrees of freedom are carried by $\hat{m}_{AB}$, while information about the background on which we compactify is contained in the twist matrices. We can adopt the point of view that we are only interested in studying properties of this background, in which case we take $\hat{m}_{AB}$ to be constant and identify the twist matrices with the generalised vielbein for the background:
\be
\hat{m}_{AB} ( \mathbb{X} ) \rightarrow \delta_{\alpha \beta} \quad , \quad W^A{}_a (x)\rightarrow E^{\alpha}{}_a (x) \,.
\ee
The situation thus reduces to that which we have been studying so far in this paper. We shall now continue in this framework, and not explicitly refer to the gauged Scherk-Schwarz setting again: however, we will remember that we have these close links. In particular, although we will not study this in this paper, the quadratic constraints resulting from this setting may be used to derive Bianchi identities for the geometric and non-geometric fluxes which we now intend to study.

\section{The torsion as generalised fluxes: M-theory fluxes}

Having found a geometrical origin for the action of the $\SLf$ extended field theory, we now want to explore the meaning of the generalised torsion from the point of view of the physical spacetime. To do so, we choose a general parametrisation of the generalised vielbein and work out the components of the generalised torsion in this parametrisation. We will be able to identify a set of spacetime tensors which appear naturally and which represent different fluxes in the spacetime picture. Some of these fluxes can be immediately interpreted geometrically, while others must be thought of as being non-geometric.

\subsection{Parametrisation and field transformations} 
\label{seMParam}

The guiding principle in writing down a parametrisation of the generalised vielbein is compatibility with the symmetries encoded in the generalised Lie derivative. For the M-theory section, we may take the following general choice (which can also be constructed as a non-linear realisation of $\SLf$ as explained in \cite{Malek:2012pw}):
\be
E^\alpha{}_a = e^{-\phi/4} 
\begin{pmatrix}
 e^{-1/2} \left( e^\mu{}_i + V^\mu W_i \right)  & e^{1/2} V^\mu \\
e^{-1/2} W_i & e^{1/2}
\end{pmatrix} \,,
\label{eq:MTheoryE}
\ee
which has inverse
\be
E_\alpha{}^a = e^{\phi/4} 
\begin{pmatrix}
e^{1/2} e_\mu{}^i & -e^{-1/2} W_\mu \\
 - e^{1/2} V^i  &  e^{-1/2} \left( 1 + V^j W_j \right)
\end{pmatrix} \,.
\ee
The fields appearing here are as follows. We have a spacetime vielbein $e^{\mu}{}_i$ with
determinant $e \equiv | \det e |$, and the scalar $\phi$ coming from the truncation (explicitly, one
should take $e^{\phi} = |g_7|^{1/7}$, where $g_7$ is the determinant of the metric in the external
directions). The vector $V^i$ is a dualisation of the three-form:
\be
V^i = \frac{1}{3!} \epsilon^{ijkl} C_{jkl} \,,
\ee 
and the covector $W_i$ is a dualisation of an antisymmetric field with three-vector indices:
\be
W_i = \frac{1}{3!} \epsilon_{ijkl} \Omega^{jkl} \,.
\ee
We refer to this as a dual field.\footnote{Parameterisations of the generalised vielbein using a dual field were considered for string theory in \cite{Grana:2008yw} and \cite{Andriot:2011uh}. In \cite{Andriot:2011uh} the parameterisation in terms of a dual field is interpreted as a field redefinition of the supergravity variables.}

Ordinarily one uses the local $\mathrm{SO}(5)$ symmetry of the generalised vielbein to remove the dual field.\footnote{Note that this local group has $10$ components. Six of these are an $\mathrm{SO}(4)$ used to ensure the spacetime metric has 10 rather than 16 components, leaving a remaining 4 to set $\Omega^{ijk}$ to zero.} However, in non-geometric situations (and also in certain cases when one has timelike directions \cite{Malek:2013sp}) the local transformation needed to remove $\Omega^{ijk}$ turns out to not be globally well-defined. This is discussed in the context of string theory in \cite{Grana:2008yw}. In order to take into account all possible situations and parametrisations, we therefore include this field.\footnote{In DFT, a similar general parameterisation of the vielbein was used to describe geometric and non-geometric fluxes of the electric sector of half-maximal gauged SUGRA \cite{Aldazabal:2011nj,Andriot:2012an,Geissbuhler:2013uka}.} 

Although we appear to have both $C_{ijk}$ and $\Omega^{ijk}$ present, this does not mean we have introduced additional degrees of freedom. The local $\mathrm{SO}(5)$ symmetry is instead unbroken and can be used to relate different configurations. However, only $\mathrm{SO}(5)$ invariant combinations appear in the Lagrangian \eqref{eq:SLft2Lag}. Thus the Lagrangian only contains specific, $\mathrm{SO(5)}$ invariant, combinations of $g_{ij}$, $C_{ijk}$ and $\Omega^{ijk}$. Note that this is why we do not have to impose some constraint involving the physical field and its dual, as was proposed in \cite{Andriot:2013xca} in the NS-NS sector of type II, to remove extraneous degrees of freedom.

We can decompose the generalised Lie derivative of the generalised vielbein into components to check that \eqref{eq:MTheoryE} is a sensible parametrisation with respect to the usual splitting of the diffeomorphism parameter, $\U^{ab} \rightarrow \xi^i, \lambda^{ij}$. The vector parameter $\xi^i$ generates spacetime diffeomorphisms, while $\tilde{\lambda}_{ij} \equiv \frac{1}{2} \eta_{ijkl} \lambda^{kl}$ gives gauge transformations of the three-form. If we do not impose the section condition, the usual physical transformations will be modified by terms involving derivatives along the dual directions. 

Our goal is to use the spacetime symmetries to classify the objects appearing in our torsion
irreducibles. The natural symmetries to use are spacetime diffeomorphisms, generated by $\xi^i$.
Using the
generalised Lie derivative we find that under these transformations we have
\be
\begin{split} 
\delta_\xi e^\mu{}_i & = \mathcal{L}_\xi e^\mu{}_i \,,\\
\delta_\xi C_{ijk} & = \mathcal{L}_\xi C_{ijk} \,, \\
\delta_\xi \Omega^{ijk} & = \mathcal{L}_\xi \Omega^{ijk} - 3 \partial^{[ij} \xi^{k]} \,, \\
\end{split}
\label{eq:SpacetimeDiffeoM}
\ee
where $\mathcal{L}_\xi$ here denotes the usual spacetime Lie derivative. We see that the dual field has an unusual transformation under diffeomorphisms. This reflects the fact that it is associated to non-geometric configurations, and does not fit naturally into the usual choice of section. By choosing a different section, dual to the original, a subsector of the diffeomorphism parameters would be reinterpreted as gauge transformations of the three-form, in which case the above expression is natural. This is reminiscent of the NS-NS sector of 10-D supergravity \cite{Andriot:2012an,Andriot:2012wx}.

\subsection{Spacetime geometry}

\subsubsection{Derivatives}

In the following discussion of fluxes we will include possibly dependence on winding coordinates as this will allow us to discuss locally non-geometric configurations. However, it is important to stress that dependence on winding coordinates does not imply violation of the section condition. Indeed, we will impose the section condition throughout. In double and exceptional field theory, one can have configurations in which the fields depend on dual coordinates, which may be related to usual physical frames by acting with ``generalised duality transformations'' along non-isometry directions \cite{Shelton:2005cf,Shelton:2006fd,Dabholkar:2005ve,Hull:2009sg}. Including the winding derivatives will also allow for the possibility of off-section contributions to the fluxes (in a constrained Scherk-Schwarz setting) although we do not discuss this further here.

The partial derivative $\partial_{ab}$ on the extended space decomposes into what we interpret as the usual spatial derivative, $\partial_i$, and the antisymmetric derivatives $\partial_{ij}$. Note that the natural winding coordinates of the theory are $x_{ij}$ with lower indices: in the generalised coordinate $x^{ab}$ these are dualised using the alternating symbol $\eta^{ijkl}$, so that $x^{ij} \equiv \frac{1}{2} \eta^{ijkl} x_{kl}$. The derivative $\partial_{ij}$ is with respect to the dualised coordinate, and so actually carries a non-zero weight under spacetime diffeomorphisms.

Natural derivatives to use in the flux formulation are provided by flattening the indices on $\partial_{ab}$ using the generalised vielbein, giving the flat derivatives
\be
\mathcal{D}_{\alpha \beta} \equiv E_\alpha{}^a E_\beta{}^b \partial_{ab}\,.
\label{eq:FlatDer}
\ee
We can obtain useful combinations of derivatives by unflattening these with the \emph{spacetime} vielbein. This defines
\be
\hat\partial_{ij} \equiv e^{-\phi/2} e^{\mu}{}_i e^{\nu}{}_j \mathcal{D}_{\mu \nu} \quad , \quad
\hat{\partial}_i \equiv  e^{-\phi/2} e^{\mu}{}_i \mathcal{D}_{\mu} \,.
\ee
The additional factor of $e^{-\phi/2}$ is inserted here by hand to cancel the factor of $e^{\phi/2}$ which results from the generalised vielbein.

In terms of the ordinary spacetime and winding derivatives, we have
\be
\hat\partial_i = (1 + V^j W_j ) \partial_i - W_i V^j \partial_j - e V^j \partial_{ij} \,,
\ee
\be
\hat\partial_{ij} = e \partial_{ij} + 2 W_{[i} \partial_{j]} \,.
\ee
Note that
\be
\hat\partial_i - V^j \hat\partial_{ji} = \partial_i \,.
\label{eq:MTheoryBarPartial}
\ee
The derivative $\hat{\partial}_{ij}$ may be dualised using the alternating tensor to define a natural duality covariant extension of the winding derivatives:
\be
\hat{\partial}^{ij} \equiv \frac{1}{2} \epsilon^{ijkl} \hat{\partial}_{kl} = \partial^{ij} + \Omega^{ijk} \partial_k \,.
\ee
This is an improvement over the bare $\partial^{ij}$ derivative in the following sense\footnote{See \cite{Andriot:2012an,Andriot:2012wx} for a similar discussion for the NS-NS sector of type II supergravity}. Consider some spacetime diffeomorphism scalar, $\varphi$. Then, although $\partial_i \varphi$ is automatically a tensor, $\partial^{ij} \varphi$ is not. However, one can check that $\hat{\partial}^{ij} \varphi$ defines a spacetime tensor:
\be
\delta_\xi \hat{\partial}^{ij} \varphi = \mathcal{L}_\xi \hat{\partial}^{ij} \varphi \,,
\ee
up to the section condition. Note that the latter is obeyed by $\hat{\partial}^{ij}$ and $\partial_i$, i.e. we have
\be
\hat{\partial}^{ij} f \partial_i g + \partial_i f \hat{\partial}^{ij} g = 0 
\quad ,
\quad
\hat{\partial}^{[ij} f \hat{\partial}^{kl]} g = 0 \,.
\ee
Although we are not explicitly solving the section condition in the sense of setting $\partial^{ij} = 0$ everywhere, we still impose the section condition as a constraint.

\subsubsection{Tensors}

In order to build tensors under spacetime diffeomorphisms, we first introduce flat connections for both types of derivatives:
\be
\Gamma_{ij}{}^k \equiv e_\mu{}^k \partial_i e^{\mu}{}_j \,,
\ee
\be
\hat{\Gamma}^{ij}{}_k{}^l \equiv e_\mu{}^l \hat\partial^{ij} e^{\mu}{}_k \,,
\ee
with associated covariant derivatives, $\nabla_i$ and $\hat{\nabla}^{ij}$:
\be
\nabla_{i} \varphi^k \equiv \partial_i \varphi^k + \Gamma_{i j}{}^k \varphi^j \,,
\ee
\be
\hat{\nabla}^{ij} \varphi^l = \hat\partial^{ij} \varphi^l + \hat\Gamma^{ij}{}_k{}^l \varphi^k \,.
\ee
 Under spacetime diffeomorphisms we have
\be
\delta_\xi \Gamma_{ij}{}^k = \mathcal{L}_\xi \Gamma_{ij}{}^k + \partial_i \partial_j \xi^k \,,
\ee
\be
\delta_\xi \hat{\Gamma}^{ij}{}_k{}^l = \mathcal{L}_\xi \hat{\Gamma}^{ij}{}_k{}^l + \hat{\partial}^{ij} \partial_k \xi^l \,.
\label{eq:HatGammaTransf}
\ee
These connections can be used to construct torsion-like quantities. 

Let us now list the various spacetime tensors which can be constructed from these ingredients, giving also their classification according to the decomposition to four-dimensional spacetime tensors, under $\SLf \rightarrow \mathrm{SL}(4)$. These tensors will be the geometric and non-geometric fluxes that appear in the $\SLf$ torsion.  The situation we find is quite analogous to that of the well-known $H$-, geometric, $Q$- and $R$-fluxes in string theory which were discussed in a similar fashion in \cite{Andriot:2012an}, and we therefore use similar language. We wish to stress, however, that our tensors are based on a different spacetime connection to that used in \cite{Andriot:2012an}. As a result, the tensors here will not necessarily reduce in a straightforward manner to those considered in the supergravity context \cite{Andriot:2012an} upon reducing to IIA.

\vspace{1em}
\noindent\textbf{F-flux:} The field strength of the three-form is
\be
F_{ijkl} = 4 \partial_{[i} C_{jkl]} \,.
\label{eq:MTensorF} 
\ee
This lives in the trivial representation $\mathbf{1}$ of $\mathrm{SL}(4)$.

\vspace{1em}
\noindent\textbf{Geometric flux:} 
The natural spacetime Weitzenb\"ock torsion is as usual:
\be
T_{ij}{}^k \equiv \Gamma_{ij}{}^k  - \Gamma_{ji}{}^k \,.
\label{eq:MTensorTorsion} 
\ee
This is known as geometric flux. Its trace and trace-free parts correspond to the irreducible representations $\mathbf{4}$ and $\mathbf{20}$ of $\SL{4}$. 

\vspace{1em}
\noindent\textbf{$Q$-flux:} This is a globally non-geometric flux, given by the tensor
\begin{equation}
 \mathcal{Q}_{i}{}^{jkl} \equiv Q_i{}^{jkl} + 3 \hat{\Gamma}^{[jk}{}_i{}^{l]} \,,
\end{equation}
where we defined
\begin{equation}
Q_{i}{}^{jkl} \equiv \partial_i \Omega^{jkl} \,.
\label{eq:MTensorQ1} 
\end{equation}
This is not a tensor by itself: Under a spacetime diffeomorphism, the transformation \eqref{eq:SpacetimeDiffeoM} of $\Omega^{ijk}$ leads to
\be
\delta_\xi Q_{i}{}^{jkl} = \mathcal{L}_\xi Q_{i}{}^{jkl} - 3 \hat{\partial}^{[jk} \partial_i \xi^{l]} \,.
\ee
From \eqref{eq:HatGammaTransf} one can see that the anomalous variation is cancelled by the winding connection term $3 \hat{\Gamma}^{[jk}{}_{i}{}^{l]}$.

The Q-flux, $\mathcal{Q}_i{}^{jkl}$, fits into the $\mathbf{6} \oplus \mathbf{10}$ representation of $\mathrm{SL}(4)$, corresponding again to the trace and trace-free parts. We can also define the dualised form
\be
\mathcal{Q}_{i,j} \equiv \frac{1}{3!} \epsilon_{jklm} \mathcal{Q}_{i}{}^{klm} \,,
\label{eq:MTensorQ2} 
\ee 
in which case the $\mathbf{6}$ and $\mathbf{10}$ correspond to the antisymmetric and symmetric parts. 

\vspace{1em}
\noindent\textbf{$R$-flux:} This is a locally non-geometric flux (i.e. it involves a dependence on a
dual coordinate). By acting with a hatted winding derivative on the dual field we can define a tensor
\be
R^{i,jklm} = 4 \hpartial^{i[j} \Omega^{klm]} \,.
\label{eq:MTensorR} 
\ee
This lives in a $\mathbf{4}$ of $\mathrm{SL}(4)$. The dual may be defined as
\be
L^i \equiv \frac{1}{3!} \epsilon_{jklm} \hpartial^{ij} \Omega^{klm} \,.
\label{eq:MTensorL} 
\ee

\vspace{1em}
\noindent\textbf{$\nT$-flux:} The quantity
\be
\nT^{i,j} \equiv \hat{\Gamma}^{ki}{}_k{}^j \,,
\label{eq:MTensorHatT} 
\ee
also transforms as a tensor. It lives in a $\mathbf{6} \oplus \mathbf{\bar{10}}$ of $\mathrm{SL}(4)$. 
\vskip1em

\noindent Finally, we will also have winding derivatives of the three-form:
\be
\hat{\nabla}^{ij} C_{jkl} \,,
\ee
which will turn out to usually appear in the dualised form $\hat{\nabla}_{ij} V^k$, 
\be
\hat{\nabla}_{ij} V^k = 4 \delta^k_{[i} \hat{\nabla}^{lm} C_{jlm]} \,,
\ee
giving additional pieces in the $\mathbf{4}$ and $\mathbf{20}$ of $\mathrm{SL}(4)$. This is a
spacetime diffeomorphism tensor although it is not gauge invariant.

\subsection{Decomposition of the torsion irreps}

We can now give the decomposition of the generalised torsion in terms of the above tensors (some of the intermediate results in this calculation may be found in appendix \ref{seDetails}). This will allow us to understand the effect of dualities on flux backgrounds as we will demonstrate using examples in section \ref{se:MExamples}

\vspace{1em}
\noindent\textbf{$\mathbf{15}$:}
We have
\be
\begin{split} 
S_{55} & = 4 e \nabla_k V^k =
\frac{e}{3!} \epsilon^{ijkl} F_{ijkl} - \frac{2e}{3} \epsilon^{ijkl} C_{ijk} T_{lm}{}^m \,,
\\
S_{i5} & = 2 T_{ki}{}^k + 2 \hat\nabla_{ki} V^k + e^{-1} W_i S_{55}  \,,
\\
S_{ij} & = - 4 e^{-1} \mathcal{Q}_{(i,j)} + e^{-1} 2W_{(i} S_{j)5} - e^{-2} W_i W_j S_{55}\,.
\label{eq:MSCurved}
\end{split} 
\ee
The recursive form and the factors of $e$ are required by the generalised Lie derivative. In terms of group theory, the $\mathbf{15}$ of $\SLf$ reduces to the $\mathbf{10}\oplus \mathbf{4} \oplus \mathbf{1}$ of $\mathrm{SL}(4)$. It is easy to identify these:
\be
\begin{split}
\mathbf{10} : & \quad \mathcal{Q}_{(i, j) } \,, \\ 
\mathbf{4} : & \quad \hat{\nabla}_{j i} V^j + T_{j i}{}^i \,,\\ 
\mathbf{1} : & \quad  F_{ijkl} \,. \\ 
\end{split} 
\ee

\vspace{1em}
\noindent\textbf{$\mathbf{10}$:}
We have
\be
\begin{split}
\tau_{i5} &  = \frac{1}{2} \nabla_{ik} V^k - \frac{1}{2} T_{ik}{}^k - \frac{3}{2} \partial_i \phi \,,
\\ 
\tau_{ij} & = e^{-1} \left(
\epsilon_{ijkl} \nT^{k,l} + Q_{[i,j]} - 2 W_{[i } \tau_{j]5} - \frac{3}{2} \hat{\partial}_{ij} \phi 
\right) \,.
\label{eq:MTauCurved} 
\end{split} 
\ee
Here we see the $\mathbf{10}$ of $\SLf$ reduces to the $\mathbf{4}\oplus \mathbf{6}$ of $\mathrm{SL}(4)$. 

\vspace{1em}
\noindent\textbf{$\mathbf{40}$:}
We obtain
\be
\widetilde T_{ij5}{}^k = -T_{ij}{}^k - \frac{2}{3} \delta^k_{[i} T_{j]l}{}^l +  \nabla_{ij} V^k + \frac{2}{3} \delta^k_{[i} \nabla_{j]l} V^l \,,
\label{eq:MTCurvedij5k}
\ee
\be
\widetilde T_{ijk}{}^l =  e^{-1} \epsilon_{ijkm} \nT^{m,l} +  2 e^{-1}\delta^{l}_{[i} \epsilon_{jk]mn} \nT^{m,n} 
+ 2 e^{-1} \mathcal{Q}_{[i,j} \delta_{k]}^{l} + 3 e^{-1} W_{[i} \widetilde  T_{jk]5}{}^l \,,
\label{eq:MTCurvedijkl}
\ee
\be
\widetilde T_{ij5}{}^5 = - \frac{4}{3}e^{-1} \mathcal{Q}_{[i,j]} - \frac{1}{3} e^{-1} \epsilon_{ijkl} \nT^{k,l} - e^{-1} W_k \widetilde T_{ij5}{}^k \,,
\label{eq:MTCurvedij55}
\ee
\be
\widetilde T_{ijk}{}^5 = -  e^{-2} \epsilon_{ijkl} L^l- e^{-1} W_l \widetilde T_{ijk}{}^l 
+ 3 e^{-1} W_{[i} \widetilde T_{jk]5}{}^5 
-3 e^{-2} W_l W_{[i} \widetilde T_{jk]5}{}^l \,.
\label{eq:MTCurvedijk5}
\ee
Observe that these are not automatically spacetime irreducible representations: we have
$\widetilde{T}_{i j k}{}^k = - \widetilde{T}_{i j 5}{}^5$ as a consequence of the
tracelessness of $\widetilde{T}$. Let us dualise the former,
\be
\widetilde{T}^{i , j } \equiv \frac{1}{3!}\epsilon^{i k l m} \widetilde{T}_{klm}{}^j 
 = 
\frac{1}{3} e^{-1} \epsilon^{i j k l} \mathcal{Q}_{k,l} 
+ \frac{1}{3} e^{-1} \nT^{[i,j]} -e^{-1}\nT^{(i,j)} 
+ \frac{1}{2} e^{-1} \epsilon^{iklm} W_k \tilde{T}_{lm5}{}^j \,.
\ee 
We can then check we have $\widetilde{T}^{[i,j]} = - \frac{1}{4}
\epsilon^{ijkl} \widetilde{T}_{kl 5}{}^5$. Hence the true $\mathrm{SL}(4)$
irreducibles may be identified as 
\be
\widetilde{T}^{[i,j]} =\frac{1}{3} e^{-1} \epsilon^{i j k l} \mathcal{Q}_{k,l} + \frac{1}{3} e^{-1} \nT^{[i,j]}
+ \frac{1}{2} e^{-1} \epsilon^{[i| klm} W_k \widetilde{T}_{lm5}{}^{|j]}
\ee
\be
\widetilde{T}^{(i,j)} =  -e^{-1}\nT^{(i,j)}
+ \frac{1}{2} e^{-1} \epsilon^{(i| klm} W_k \widetilde{T}_{lm5}{}^{|j)}
\ee 
The $\mathbf{40}$ here decomposes as $\mathbf{40} \rightarrow \mathbf{\bar{4}} \oplus \mathbf{6} \oplus \mathbf{\bar{10}} \oplus \mathbf{20}$ and we can identify
\be
\begin{split}
\mathbf{20} : & \quad  \hat\nabla_{ij} V^k - T_{ij}{}^k - \mathrm{trace} \,, \\ 
\mathbf{\bar{10}} : & \quad \nT^{(i,j)}  \,, \\ 
\mathbf{6} : & \quad \mathcal{Q}_{[i, j] } \,, \\ 
\mathbf{\bar{4}} : & \quad  L^i  \,. \\ 
\end{split} 
\ee

\section{M-theory flux examples}
\label{se:MExamples}

In this section, we wish to present some examples of easily obtainable non-geometric backgrounds in string theory and M-theory which are best described in the framework of an extended theory. We will focus here on backgrounds which can be obtained by dualising a geometric background with a single flux.
Although we will not be presenting novel solutions, we wish to stress the point that the approach of this paper allows one to fully understand the non-geometric fluxes that appear.

\subsection{The string theory prototype} 
\label{se:NSNStoy}

First, let us describe the well-known prototypical toy example for the NS-NS sector \cite{Kachru:2002sk,Andriot:2011uh}.
As usual, we will start with a flat 3-torus with $H$-flux:
\begin{equation}
 \begin{split}
  ds^2 &= dx^2 + dy^2 + dz^2 \,, \\
  B_2 &= Nz dx \wedge dy\,.
 \end{split}
\end{equation}
The $H$-flux is $H_{xyz} = N$.

Dualising along the $x$-direction one obtains a twisted torus:
\begin{equation}
 \begin{split}
  ds^2 &= \left( d\tilde{x} - N z dy \right)^2 + dy^2 + dz^2 \,, \\
  B_2 &= 0 \,.
 \end{split}
\end{equation}
The geometric flux of this background
\begin{equation}
 T_{ij}{}^k = e_\mu{}^k \partial_{[i} e^{\mu}{}_{j]} \,,
\end{equation}
is non-zero: $T_{yz}{}^{\tilde{x}} = N$.

Another duality, this time along the $y$-direction, gives a globally non-geometric background with
$Q$-flux. The usual metric and Kalb-Ramond form are then globally ill-defined
\begin{equation}
 \begin{split}
  ds^2 &= \frac{d\tilde{x}^2 + d\tilde{y}^2}{1 + N^2 z^2} + dz^2 \,, \\
  B_2 &= - \frac{Nz}{1 + N^2 z^2} d\tilde{x} \wedge d\tilde{y} \,.
 \end{split}
\end{equation}
This is because the local $\mathrm{SO}(5)$ transformation that would be needed to remove the $\beta^{ij}$ field in the generalised vielbein is globally ill-defined. However, one could instead remove the $B$-field. In the resulting ``non-geometric'' frame the background is
\begin{equation}
 \begin{split}
  d{s}^2 &= d\tilde{x}^2 + d\tilde{y}^2 + dz^2 \,, \\
  \beta^2 &= N z \partial_{\tilde{x}} \wedge \partial_{\tilde{y}} \,.
 \end{split}
\end{equation}
This non-geometric background has a $Q$-flux
\begin{equation}
 Q^{\tilde{x}\tilde{y}}{}_z = \partial_z \beta^{\tilde{x}\tilde{y}} = N \,.
\end{equation}

Finally, one can perform a duality along the $z$-direction, which is not an isometry, to obtain a
locally non-geometric background,
\begin{equation}
 \begin{split}
 d{s}^2 &= d\tilde{x}^2 + d\tilde{y}^2 + d\tilde{z}^2 \,, \\
 \beta^2 &= N  z \partial_{\tilde{x}} \wedge \partial_{\tilde{y}} \,.
 \end{split}
\label{eq:NSNSRsoln}
\end{equation}
This background depends explicitly on $z$, which in this frame is a dual coordinate. Hence we say
that there is no local geometric description. The $R$-flux of this background is
\begin{equation}
 R^{\tilde{x}\tilde{y}\tilde{z}} = 3\partial^{[\tilde{x}} \beta^{\tilde{y}\tilde{z}]} = N \,.
\end{equation}
This chain of dualities is summarised by saying that
\begin{equation}
 H_{xyz} \rightarrow T^x{}_{yz} \rightarrow Q^{xy}{}_z \rightarrow R^{xyz} \,,
\end{equation}
Thus, we see that a single Buscher T-duality lifts an index from a subscript to a superscript \cite{Shelton:2005cf}. This is best understood as the action of T-duality on the $\ODD$ generalised torsion of the Weitzenb\"ock connection \cite{Berman:2013uda} which analogously to the torsion considered here is a covariant $\ODD$ tensor containing the fluxes \cite{Geissbuhler:2011mx, Geissbuhler:2013uka}.

\subsection{Duality chains and an M-theory toy model} 
\label{se:MExamplesChain}

We described the fluxes of M-theory in terms of U-duality tensors. Thus, we can now find the action of U-dualities on fluxes simply by performing matrix multiplication. In order to describe duality chains similar to the above, we need to use the M-theory versions of Buscher dualities. As the M-theory U-duality groups reduce only to the T-duality subgroup $SO(D,D)$ one such U-duality can be thought of as corresponding to a pair of Buscher dualities. In fact one finds that the form of the duality in fact exchanges three directions with dual coordinates - reducing to string theory on one of these directions one is able to show that the duality descends to a Buscher duality acting on the other two (plus an exchange of coordinates) \cite{Malek:2012pw}.

The $\SLf$ element in question is 
\begin{equation}
 U^a{}_b = \begin{pmatrix}
  \delta^i{}_j - n^i \bar{n}_j & n^i \,, \\
  - \bar{n}_j & 0
 \end{pmatrix} \,,
\end{equation} 
where $n^i \bar{n}_i = 1$. 

The choice of vector $n^i$ specifies the directions in which the duality acts. Suppose our physical
coordinates are $x,y,z,w$,\footnote{We will take $\eta^{xyzw} = \eta_{xyzw} = +1$.} which we will think of as parametrising some four-torus in the examples below. Let the duality be along the $x$, $y$, $w$ directions (so that if we reduce from M-theory to string theory on the $w$ direction this descends to a usual pair of Buscher dualities on the $x$ and $y$ directions). Then we should take $n^z = \bar{n}_z = 1$, and the effect of the duality on a generalised tensor is to swap the $z$ index for a $5$ index and a $5$ index for a $z$ index, up to a sign: if $\tilde V^a = U^a{}_b V^b$ then letting $\alpha = x,y,w$ one has $\tilde V^\alpha = V^\alpha$, $\tilde V^z = V^5$, $\tilde V^5 = - V^z$.
(Similarly for $\tilde V_a = V_b (U^{-1})^b{}_a$ one has $\tilde V_\alpha = V_\alpha, \tilde V_z = V_5, \tilde V_5 = - V_z$.)

For the generalised coordinates this means that for $\tilde x^{ab} = U^a{}_c U^b{}_d x^{cd}$, 
\be
\tilde x^{\alpha \beta} = x^{\alpha \beta} \quad , \quad \tilde x^{\alpha z} = x^{\alpha 5} 
\quad , \quad 
\tilde x^{\alpha 5} = x^{z \alpha} \quad , \quad \tilde x^{z5} = x^{z5} \,.
\label{eq:MCoordBuscher}
\ee
The physical coordinates in the new frame are $\tilde x^{\alpha 5}$ and $\tilde x^{z5}$. 
We shall denote a Buscher duality along the three directions $x,y,w$ by $\Uxyw$.
 
Let us now turn to the fluxes to see what kind of non-geometric backgrounds we can obtain by dualising geometric ones. This is a much more delicate matter than for string theory because we always have to dualise along three directions. 
For simplicity, we will focus here on geometric backgrounds with just one flux, either the four-form flux or the geometric flux. 

If we start with a four-form flux turned on, then referring to the expressions \eqref{eq:MSCurved}, \eqref{eq:MTauCurved} and \eqref{eq:MTCurvedij5k} to \eqref{eq:MTCurvedijk5} for the irreducible components, we see we only have non-zero $S_{55}$. By acting with the transformation matrix $\Uxyw$ (any choice of directions could be made here) we find this can only be dualised into a $Q$-flux, corresponding to having non-zero $S_{zz}$ component: 
\begin{equation}
   F_{wxyz} \,\, \stackrel{\Uxyw}{\longleftrightarrow} \,\, \mathcal{Q}_{z,z} \equiv \mathcal{Q}_{z}{}^{wxy} \,.
\end{equation}
(Recall that the globally non-geometric $Q$-flux $\mathcal{Q}_i{}^{jkl}$, defined in \eqref{eq:MTensorQ1}, appears as the trace-free part, $\mathcal{Q}_{(i,j)}$, and also a trace part $\mathcal{Q}_{[i,j]}$.)

If instead we begin with a geometric flux of the form $T_{ix}{}^i$ and no three-form, corresponding to the torsion irreducible $S_{x5}$, then 
\begin{equation}
 T_{ix}{}^i \,\, \stackrel{\Uyzw}{\longleftrightarrow} \,\, T_{ix}{}^i \,\, \stackrel{\Uxyw}{\longleftrightarrow} \, \, \mathcal{Q}_{(x,z)} \,.
\end{equation}
Note that in this case the initial compactification is on a non-uni-modular Lie group and so we do not expect the lower-dimensional supergravity to have a consistent action principle \cite{Bergshoeff:2003ri}.

Now let us consider the other kind of geometric background:  one with traceless geometric flux, e.g. $T_{yz}{}^x$. This corresponds to the torsion irreducible $\widetilde{T}_{yz5}{}^x$ and referring to the component decompositions of this irreducible, equations \eqref{eq:MTCurvedij5k} to \eqref{eq:MTCurvedijk5}, we now find two different ways to obtain an $R$-flux:
\begin{equation}
 \begin{split}
  T_{yz}{}^x &\,\,\stackrel{\Uyzw}{\longleftrightarrow}\,\, R^{w[x,yzw]} \,, \\
  T_{xy}{}^z &\,\,\xleftrightarrow{\Uxyz}\,\, \nT^{x,x} \,\,\stackrel{\Uyzw}{\longleftrightarrow}\,\, R^{x[x,yzw]} \,. \\
 \end{split}
\end{equation}
Note that these will involve carrying out dualities along directions which are not isometries. This is of course expected to be the case for a background carrying locally non-geometric $R$-flux, and is possible within the framework of the extended theory. 

This configuration, $T_{yz}{}^x \neq 0$, is also self-dual under $\Uxzw$ or $\Uxyw$
\begin{equation}
  T_{xy}{}^z \stackrel{\Uxzw}{\longleftrightarrow} T_{xy}{}^z \stackrel{\Uxyw}{\longleftrightarrow} T_{xy}{}^z \,. \label{eq:Tselfdual}
\end{equation}
Obviously other duality chains will be possible involving more complicated set-ups. We will finish this subsection by considering a toy model that presents in detail the generalisation to M-theory of the string theory three-torus with $H$-flux. We will realise two of the above-mentioned example duality chains explicitly: $F_{wxyz} \longleftrightarrow \mathcal{Q}_{z,z}$ and $T_{yz}{}^x \longleftrightarrow R^{w[x,yzw]}$. 

We thus introduce a four-torus with coordinate $x,y,z,w$, and include a general external metric in the other seven directions as it will transform under dualities too. To be precise, the external metric will transform conformally, with the scaling determined by the transformation of the extra scalar $e^\phi$ in the generalised metric, given the identification $e^\phi = |g_7|^{1/7}$. 

For $F_{wxyz} \longleftrightarrow \mathcal{Q}_{z,z}$, our initial $T^4$ is flat and we choose a three-form with constant flux through this torus:
\be
\begin{split} 
ds^2 &= ds_7^2 + dz^2 + dw^2 + dx^2 + dy^2 \,,
\\ C_3  & = N z dw \wedge dx \wedge dy \,,
\end{split} 
\ee
This corresponds to $S_{55} = 4N$. 
We can now carry out a Buscher transformation along the $w,x,y$ directions. We find the resulting configuration to be
\be
\begin{split} 
ds^2 & = (1+N^2 z^2)^{1/3} ds_7^2 
+ (1 + N^2 z^2)^{1/3} dz^2 + (1 + N^2 z^2)^{-2/3} (d\tilde w^2 + d \tilde x^2 + d \tilde y^2) \,, \\
C_3 & = - \frac{Nz}{1+N^2 z^2} d\tilde w \wedge d \tilde x \wedge d\tilde y \,.
\end{split} 
\ee
This background is non-geometric: when using the 3-form $C_3$ to express the solution looks ill-defined globally. It needs to be patched by a U-duality transformation which is not a diffeomorphism or gauge symmetry of $C_3$. This bad behaviour is introduced because the local $\so{5}$ transformation which is needed to obtain the frame involving $C_3$ is globally ill-defined. Instead we should consider an alternative frame, containing a trivector. Using the expressions \eqref{eq:MChangeOfFrame} we get 
\be
\begin{split} 
d \tilde s^2 & = ds_7^2 + dz^2 + d\tilde w^2 + d\tilde x^2 + d\tilde y^2 \,, \\
\Omega^3 & = - N z \partial_{ \tilde x} \wedge \partial_{ \tilde y} \wedge \partial_{ \tilde w} \,.
\end{split}
\ee
In this dual frame the solution is periodic but involves a dual field. It is easy to see that there is non-zero $Q$-flux, $\mathcal{Q}_{ z}{}^{\tilde x \tilde y \tilde w} = \mathcal{Q}_{z,z} = - N$ as expected from the duality chain: we obtain $S_{zz} = 4N$ exactly as predicted by the transformation of the torsion under duality.

For the other duality chain, $T_{yz}{}^x \longleftrightarrow R^{w[x,yzw]}$, let us instead start with a twisted torus background
\be
\begin{split} 
ds^2 & =  ds_7^2 
+  dw^2 + dz^2 + ( dx - Nz dy )^2 + dy^2 \,, \\
C_3 & = 0 \,.
\end{split} 
\ee
This background has non-zero geometric flux, $T_{yz}{}^x = N$, which corresponds to the irreducible $\tilde{T}_{yz5}{}^x = - N$. This configuration is self-dual under Buscher
duality on $x,y,w$ directions as seen from \eqref{eq:Tselfdual}. Let us instead consider a Buscher
duality acting on $y,z,w$ directions. In the $C_3$ frame we have
\begin{equation}
 \begin{split}
  ds^2 &= ( 1 + N^2 z^2)^{1/3} ds_7^2 + (1 + N^2 z^2)^{1/3} d\tilde y^2 
+ ( 1 + N^2z^2)^{-2/3} \left( dx^2 + d\tilde z^2 + d\tilde w^2 \right) \\
  C_3 &= - \frac{Nz}{1+N^2z^2} d x \wedge d\tilde z \wedge d\tilde w  \,.
 \end{split}
\end{equation}
This can be seen to depend on what is now a dual coordinate, $z$, and so is not even locally
geometric. However, one can still pass to a more appropriate description with the trivector:
\be
\begin{split} 
d\tilde{s}^2 & =  ds_7^2 
+  d \tilde z^2 + d \tilde w^2 + dx^2 + d\tilde y^2 \,, \\
\Omega^3  & =  -N z \partial_{\tilde x}  \wedge \partial_{\tilde z} \wedge \partial_{\tilde w} \,.
\end{split} 
\ee
We can do nothing about the dependence on $z$, but this frame leads to a well-defined flux.
By carefully referring to the transformations \eqref{eq:MCoordBuscher}, we see we can identify $z$
with the winding coordinate $\tilde x^{zx}$, so that we have $\partial^{\tilde y \tilde w}
\Omega^{x\tilde z \tilde w} = - N$. Using the definition \eqref{eq:MTensorR} we see the $R$-flux is
\be
R^{\tilde w , x \tilde y \tilde z \tilde w} = -N
\ee
and as a result we indeed have from the decomposition \eqref{eq:MTCurvedijk5} that $\widetilde{T}_{x \tilde y \tilde z}{}^5 =  N$. 

\subsection{The $5^3$ solution} 

We will now demonstrate that our duality chains are also applicable to solutions of M-theory. We thus consider acting with dualities on M-theory solutions with similar properties to the toy examples just discussed.
One such solution is the $5^3$ brane \cite{ LozanoTellechea:2000mc, deBoer:2012ma
}. This is obtained by acting dualising the M5 brane. The solution for the latter is
\be
\begin{split}
ds^2 & = H^{-1/3} ( - dt^2 + d\vec{y}_5{}^2) + H^{2/3} d\vec{z}_{5}{}^2 \, , 
\\
C_6 & =  ( H^{-1} - 1) dt \wedge dy^1 \wedge \dots \wedge dy^5 \,,
\end{split}
\ee
where $H = 1 + \frac{k}{r^3}$ and $r\equiv |\vec{z}_5|$. We wrap the solution on a transverse $T^3$, in the $z_3,z_4,z_5$ directions and smear it in those directions. The resulting solution can then be dualised along these directions. Prior to dualising, we have
\be
\begin{split} 
ds^2 & = H^{-1/3} \left( - dt^2 + d\vec{y}_5{}^2\right) + H^{2/3} \left( dr^2 + r^2 d\theta^2 \right) 
+ H^{2/3} \left( (dz_3)^2 + (dz_4)^2 + (dz_5)^2 \right) \,,
\\
C_3 & = \sigma \theta dz_3 \wedge dz_4 \wedge dz_5 \,,
\end{split}
\label{eq:M5Wrapped}
\ee
where now $H = h_0 + \sigma \log \frac{\mu}{r}$, with constant $\sigma \equiv \frac{2k}{\pi^2 R_3 R_4 R_5}$, $\mu$ a regularisation scale and $h_0$ a divergent bare quantity (see the discussion in \cite{deBoer:2012ma} for the very similar case of the $5_2^2$ brane in string theory). We have switched to polar coordinates, $r$, $\theta$, in the $z_1,z_2$ directions. Note that the solution carries a constant $F_4$ flux. 

We now consider U-duality acting in the $z_3,z_4,z_5$ directions. The transformed solution has the form
\be
\begin{split} 
ds^2 & = H^{-1/3} K^{1/3} \left( - dt^2 + d\vec{y}_5{}^2\right) +
H^{2/3} K^{1/3} \left( dr^2 + r^2 d\theta^2\right) 
+ H^{2/3} K^{-2/3} \left( (d\tilde z_3)^2 + (d\tilde z_4)^2 + (d\tilde z_5)^2 \right) \,,
\\
\quad C_3 & = - K^{-1} \sigma \theta d \tilde z_3 \wedge d \tilde z_4 \wedge d \tilde z_5 \,,
\end{split}
\ee
where
\be
K = H^2 + \sigma^2 \theta^2 \,.
\ee
This is a non-geometric solution: it is not single-valued for $\theta \rightarrow \theta + 2\pi$, even modulo coordinate transformations and gauge transformations. However, it can be seen to transform by a duality transformation as $\theta \rightarrow \theta + 2 \pi$. The solution is thus an example of a U-fold. This is the M-theory analogue of the ``$Q$-brane'' in string theory \cite{Hassler:2013wsa} and is also known as the $ 5_3$ brane.

The fact that we have such unpleasant behaviour of our physical fields is a consequence of using an unsuitable parametrisation. We should as before instead use a non-geometric frame, exchanging the three-form for a trivector $\Omega^3$. Carrying out the field redefinition using the generalised metric \eqref{eq:MChangeOfFrame}, one obtains the new form of the solution:
\be
\begin{split} 
ds^2 & = H^{1/3} \left( - dt^2 + d\vec{y}_5{}^2\right) +
H^{4/3} \left( dr^2 + r^2 d\theta^2\right) 
+ H^{-2/3}  \left( (d\tilde z_3)^2 + (d\tilde z_4)^2 + (d\tilde z_5)^2 \right) \,,
\\
\quad \Omega^{345} & = - \sigma \theta  \,.
\end{split}
\ee
We see now that this solution is well-defined for $\theta \rightarrow \theta + 2\pi$, up to a simple gauge transformation of the trivector. Such a transformation has no simple interpretation in terms of the usual geometric and physical variables, and is the source of the non-geometric behaviour. It has constant M-theoretic $Q$-flux, $\mathcal{Q}_{\theta}{}^{345} = - \sigma$. 

Similarly, one could start with the M-theory Kaluza-Klein monopole, which carries geometric flux, and carry out a duality transformation along a non-isometry direction to reach a configuration with $R$-flux, the analogue of the ``$R$-brane'' in string theory \cite{Hassler:2013wsa}.

\section{The torsion as generalised fluxes: IIB fluxes}

We now repeat the analysis of the previous sections for the case where we choose a parametrisation of the generalised vielbein that, after choosing an inequivalent section choice, leads to IIB supergravity \cite{Blair:2013gqa}.

\subsection{Parametrisation and field transformations} 

For IIB, by noticing that the little metric in M-theory parametrisation has a similar form to the inverse little metric in IIB parametrisation, we may take
\be
E^\alpha{}_a = 
e^{-\phi/4}
\begin{pmatrix}
e^{1/2} e^{\bmu}{}_\mu & e^{-1/2} W^{\bmu}_i \\
e^{1/2} V^{\bi}_\mu & e^{-1/2} \left( h^{\bi}{}_i + V^{\bi}_\rho W^\rho_{i} \right) 
\end{pmatrix}  \,,
\label{eq:IIBGenViel}
\ee
with inverse
\be
E^a{}_\alpha = e^{\phi/4} \begin{pmatrix}
e^{-1/2} \left( e^\mu{}_{\bmu} + W^\mu_k V_{\bmu}^k \right) & - e^{-1/2} W^\mu_{\bi} \\
- e^{1/2} V_{\bnu}^i & e^{1/2} h^i{}_{\bj}  
\end{pmatrix} \,.
\ee
Here $g \equiv \det (g^{\mu \nu})$, with $e_{\bmu}{}^\mu$ the vielbein for this metric. Meanwhile
$h^{\bi}{}_i$ is a vielbein for the unit determinant matrix of scalars,
$\mathcal{M}_{ij}$ (and so parametrises the coset $\mathrm{SL}(2)/\mathrm{SO}(2)$). Again we have
the scalar $\phi$ related to the truncation, with $e^\phi = |g_7|^{1/7}$.

We have that $V^i_\mu$ is a dualisation of the two two-forms, $V^i_\mu = \frac{1}{2} \epsilon_{\mu \nu \rho} B^{i \nu \rho}$, while similarly $W^\mu_i = \frac{1}{2} \epsilon^{\mu \nu \rho} \beta_{i \nu \rho}$ (here $\epsilon^{\mu \nu \rho} = g^{1/2} \eta^{\mu\nu \rho}$). The preceding involve what we take as the ``natural'' position of the $\mathrm{SL}(2)$ index in defining these objects and the bivectors $\beta^{i \mu\nu}$ include the original bivector field of 10-d supergravity \cite{Andriot:2011uh} (usually simply referred to as $\beta^{\mu \nu}$) as well as its S-dual. Note that there will be a further dual field appearing in $h^{\bi}{}_i$ (as an alternative to the Ramond-Ramond zero form). Hence we have included dual fields for all form fields appearing in the generalised vielbein. See section \ref{seMParam} for a discussion of the relationship of the local $\mathrm{SO}(5)$ symmetry and the form-potentials and their dual fields in the generalised vielbein.

In the IIB parametrisation, the coordinates $x^{ab}$ lead to physical coordinates $x_\mu \equiv
\frac{1}{2} \eta_{\mu \nu \rho} x^{\nu \rho}$, alongside dual coordinates $x^{\mu i}$ and $x^{ij}$.
The generalised diffeomorphism parameter $\U^{ab}$ vector $\xi_\mu \equiv \frac{1}{2} \eta_{\mu \nu \rho} U^{\nu \rho}$, which generates spacetime diffeomorphisms, a pair of 1-forms, $\lambda^{i\mu}$, which generate gauge transformations of the 2-forms $B^{i\mu \nu}$, and an additional component $\U^{ij}$, which vanishes from the transformation rules when the IIB section is imposed. 

Note that
$S$-duality (acting on the $\mathrm{SL}(2)$ indices $i,j$) is manifest in this parametrisation, and
as a result when the action \eqref{eqSL5action} is evaluated using \eqref{eq:IIBGenViel} with
$W^\mu_{\bi} = 0$ we reach
(a truncation of) the IIB supergravity action in Einstein frame \cite{Blair:2013gqa}.

We can evaluate the transformation properties of the fields under these transformations using the generalised Lie derivative. As before, we will focus on the classification of tensors and other objects in the theory using spacetime diffeomorphisms. Note that these are defined by
\be
\delta_\xi \varphi_\mu \equiv \mathcal{L}_\xi \varphi_\mu = \xi_\nu \partial^\nu \varphi_\mu - \varphi_\nu \partial^\nu \xi_\nu \,.
\ee
The dualisation of the coordinates means that vectors carry a lower index. 

Starting from the vielbein or generalised metric, one can show that
\be
\begin{split} 
\delta_\xi e^{\mu}{}_{\bmu} & = \mathcal{L}_\xi e^{\mu}{}_{\bmu} \,,\\
\delta_\xi B^{i \mu \nu} & = \mathcal{L}_\xi B^{i \mu \nu} \,,\\
\delta_\xi \beta_{i \mu \nu} & = \cl_\xi \beta_{i \mu \nu} + 2 \partial_{i[\mu} \xi_{\nu]} \,.
\end{split} 
\ee
Again, we see that the dual fields have an unusual transformation under spacetime diffeomorphisms, just as was noted for the NS-NS sector in \cite{Andriot:2012an,Andriot:2012wx}.

\subsection{Spacetime geometry}

\subsubsection{Derivatives}

We have the same flattened partial derivatives \eqref{eq:FlatDer} as before. We obtain useful combinations of derivatives by curving with the spacetime vielbein on flat spacetime indices, and with the scalar coset vielbein $h^{\bi}{}_i$ on flat scalar indices:
\be
\begin{split}
\hat{\partial}_{ij} & \equiv e^{-\phi/2} h^{\bi}{}_i h^{\bj}{}_j \mathcal{D}_{\bi\bj} \,, \\
\hat{\partial}_{\mu i} & \equiv e^{-\phi/2} e^{\bmu}{}_\mu h^{\bi}{}_i \mathcal{D}_{\bmu \bi} \,, \\
\hat{\partial}_{\mu \nu} & \equiv e^{-\phi/2}  e^{\bmu}{}_\mu  e^{\bnu}{}_\nu  \mathcal{D}_{\bmu \bnu} \,.
\end{split} 
\ee
In terms of the vanilla spacetime and winding derivatives,
\be
\hat{\partial}_{ij} =  e \partial_{ij} + e^{-1} W_i^\mu W_j^\nu \partial_{\mu \nu} + 2
W_{[i}^\mu \partial_{j] \mu} \,,
\ee
\be
\hat{\partial}_{\mu i} = \tilde\partial_{\mu i} - V_\mu^k \hat{\partial}_{ki} \,,
\ee
\be
\hat{\partial}_{\mu \nu} = e^{-1} \partial_{\mu \nu} + 2 V_{[\mu}^i \tilde{\partial}_{\nu] i} + V_\mu^i V_\nu^j \hat{\partial}_{ij} \,.
\ee
Here we have introduced the quantity
\be
\tilde \partial_{\mu i} = \partial_{\mu i} + \beta_{i\mu \nu} \partial^\nu = \partial_{\mu i} - e^{-1} W_i^\rho \partial_{\mu \rho} \,,
\ee
which is a generalisation of the anholonomic dual derivative introduced for the NS-NS sector of type II supergravity in \cite{Andriot:2012an,Andriot:2012wx}.

The structure is a little more intricate in this case than it was for M-theory. The derivatives we choose to express our tensors in are going to be $\partial^\mu \equiv \frac{1}{2} \eta^{\mu \nu \rho} \partial_{\nu \rho}$, $\tilde\partial_{\mu i}$, and $\hat{\partial}_{ij}$. All three of these derivatives have the property that if $\varphi$ is a scalar, then the derivative of $\varphi$ is a tensor, up to the section condition. Note that the section condition is obeyed using these derivatives.

\subsubsection{Tensors}

We introduce flat connections built out of the above derivatives:
\be
\Gamma^{\mu \nu}{}_{\rho} \equiv e^{\bmu}{}_\rho \partial^\mu e_{\bmu}{}^\nu \,,
\quad 
\tilde\Gamma_{\mu i}{}^{\nu}{}_{\rho} \equiv e^{\bmu}{}_\rho \tilde\partial_{\mu i} e_{\bmu}{}^\nu \,,
\quad 
\hat{\Gamma}_{ij}{}^{\nu}{}_{\rho} \equiv e^{\bmu}{}_\rho \hat\partial_{ij} e_{\bmu}{}^\nu \,.
\ee
Up to section condition, we have
\be
\delta_\xi \Gamma^{\mu \nu}{}_{\rho} = \mathcal{L}_\xi \Gamma^{\mu \nu}{}_{\rho} + \partial^\mu \partial^\nu \xi_\rho \,,
\ee
\be
\delta_\xi \tilde\Gamma_{\mu i}{}^{\nu}{}_{\rho} = \mathcal{L}_\xi \tilde\Gamma_{\mu i}{}^{\nu}{}_{\rho} + \tilde\partial_{\mu i} \partial^\nu \xi_\rho \,,
\label{eq:TildeGammaTransf}
\ee
\be
\delta_\xi \hat{\Gamma}_{ij}{}^{\nu}{}_{\rho} = \mathcal{L}_\xi \hat{\Gamma}_{ij}{}^{\nu}{}_{\rho} + \hat{\partial}_{ij} \partial^\nu \xi_\rho \,.
\ee
We also define `connections' (which are in fact spacetime tensors) built using the scalar
vielbein:
\be
\Gamma^{\mu}{}_i{}^j = h_{\bi}{}^j \partial^\mu h^{\bi}{}_i 
\,,\quad
\tilde \Gamma_{\mu k i}{}^j = h_{\bi}{}^j \tilde \partial_{\mu k} h^{\bi}{}_i 
\,,\quad
\hat \Gamma_{k l i}{}^j = h_{\bi}{}^j \hat\partial_{k l}  h^{\bi}{}_i 
\,.\quad
\label{eq:scalarconnection} 
\ee
Note the differing index positions in these definitions. In general, when we have an object
$\varphi^i_\mu$ carrying both a spacetime and an $S$-duality index, we have by definition
\be
\nabla_{A} \varphi^i_\mu = \partial_A \varphi^i_\mu + \Gamma_{A j}{}^i \varphi^i_\mu +
\Gamma_{A}{}^{\nu}{}_{\mu} \varphi^j_{\nu} \,,
\ee
for $A$ any index we are considering: $A = {}^\mu, {}_{\mu i}, {}_{i j}$.

We can now use these to give the full set of spacetime tensors which appear. We may classify them group theoretically according to their spacetime tensor structure and behaviour under S-duality, corresponding to the decomposition $\mathrm{SL}(5) \rightarrow \mathrm{SL}(3) \times \mathrm{SL}(2)$. Before listing the tensors we find, we wish to reiterate that our geometric construction here uses a different connection to that previously used to discuss 10-dimensional supergravity \cite{Andriot:2012an}.

\vspace{1em}
\noindent\textbf{H-fluxes:} We have a pair of S-dual field strengths,
\be
H^{i \mu \nu \rho} \equiv 3 \partial^{[\mu} B^{|i| \nu \rho ]} \,,
\ee
in the $(\mathbf{1}, \mathbf{2} )$ of $\mathrm{SL}(3) \times \mathrm{SL}(2)$.

\vspace{1em}
\noindent\textbf{Geometric flux:} The usual geometric flux is just
\be
T^{\mu \nu}{}_\rho = \Gamma^{\mu \nu}{}_{\rho}  -\Gamma^{\nu \mu}{}_{\rho} \,.
\ee
This exists in the $\mathbf{3} \oplus \mathbf{6}$ representation of $\mathrm{SL}(3)$, and is invariant under the $\mathrm{SL}(2)$ S-duality.

\vspace{1em}
\noindent\textbf{$Q$-fluxes:} We have a pair of S-dual non-geometric $Q$-fluxes, one for each dual bivector. They are defined by
\be
\mathcal{Q}^{\mu}{}_{i \nu \rho} = Q^\mu{}_{i \nu \rho } - 2 \tilde\Gamma_{ i [ \nu}{}^\mu{}_{\rho] }\,,
\ee
where
\be
Q^\mu{}_{i \nu \rho} = \partial^\mu \beta_{i \nu \rho} \,.
\ee
This is a tensor under spacetime diffeomorphisms and corresponds to a $(\mathbf{3} , \mathbf{2} ) \oplus ( \mathbf{6} , \mathbf{2} )$ of $\mathrm{SL}(3) \times \mathrm{SL}(2)$.
The first term $Q^\mu{}_{i\nu\rho}$ is not a tensor by itself: we have
\be
\delta_\xi Q^\mu{}_{i \nu \rho } = \cl_\xi Q^\mu{}_{i \nu \rho} - 2 \tilde \partial_{i [\nu}
\partial^\mu \xi_{\rho]} \,.
\ee
However, comparing with equation \eqref{eq:TildeGammaTransf} we see that the connection $\tilde{\Gamma}_{i[\nu}{}^\mu{}_{\rho]}$ cancels the anomalous variation.

\vspace{1em}
\noindent\textbf{$R$-flux:} The $R$-flux structure is somewhat involved. Consider the combination
\be
\tilde\partial_{\mu i} \beta_{j \nu \rho} \,,
\ee
for which 
\be
\delta_\xi \tilde\partial_{\mu i} \beta_{j \nu \rho} =
\cl_\xi \tilde\partial_{\mu i} \beta_{j \nu \rho} + 
 2 \tilde \partial_{\mu i} \tilde \partial_{j [\nu} \xi_{\rho]} \,,
\ee
where the derivatives on the right only act on $\xi_\rho$. It turns out that this can be completed to form two tensors,
\be
R_{ij} \equiv 
\epsilon^{\mu \nu \rho} \tilde\partial_{\mu (i} \beta_{j) \nu \rho} \,,
\label{eq:IIBROld}
\ee
which lives in the $(\mathbf{1} , \mathbf{3} )$ of $\mathrm{SL}(3) \times \mathrm{SL}(2)$, as well as
\be
R^{\mu}{}_{\nu ij} \equiv \epsilon^{\mu \kappa \lambda} \tilde\partial_{\nu [i} \beta_{j] \kappa \lambda} - \hat{\Gamma}_{ij}{}^\mu{}_\nu  + \delta^\mu_\nu \hat{\Gamma}_{ij}{}^\rho{}_\rho \,.
\label{eq:IIBRNew}
\ee
which lives in the $(\mathbf{1} , \mathbf{1} ) \oplus (\mathbf{8} , \mathbf{1} )$ of $\mathrm{SL}(3) \times \mathrm{SL}(2)$.

\vspace{1em}
\noindent\textbf{$\nT$-fluxes:}
The trace
\be
\nT_{i\mu} \equiv \tilde \Gamma_{i \nu}{}^\nu{}_{\mu}
\ee 
is also a tensor, in the $(\mathbf{3} , \mathbf{2} )$ of $\mathrm{SL}(3) \times \mathrm{SL}(2)$.

\vskip1em
\noindent
There are also winding derivatives of the usual form fields:
\be
\hat{\nabla}_{i\mu} V_\nu^j 
= \frac{1}{2} \epsilon_{\nu \rho \sigma} \hat{\nabla}_{i \mu} B^{j \rho \sigma} \,, \quad \quad 
\hat{\nabla}_{i j} V_\mu^k 
= \frac{1}{2} \epsilon_{\mu \nu \rho} \hat{\nabla}_{i j} B^{k \nu\rho}\,.
\ee
These give pieces in $(  \mathbf{\bar{3}} \oplus  \mathbf{\bar{6}} , \mathbf{1} \oplus \mathbf{3})$ and $( \mathbf{3}, \mathbf{2} )$ of $\mathrm{SL}(3) \times \mathrm{SL}(2)$, respectively. 

\vspace{1em}
\noindent\textbf{Scalar fluxes:} Finally, the definitions \eqref{eq:scalarconnection}
may be taken as providing a set of scalar fluxes for each derivative:
\be
\Gamma^{\mu}{}_{i}{}^{j} , \tilde{\Gamma}_{\mu i j}{}^{k} , \hat{\Gamma}_{ijk}{}^{l}\,.
\label{eq:IIBScalarFlux}
\ee
These are tensors in the $(\mathbf{3} , \mathbf{3} )$, $(\mathbf{3} , \mathbf{2} \oplus \mathbf{4})$ and the $(\mathbf{1} , \mathbf{3} )$ representations of $\mathrm{SL}(3) \times \mathrm{SL}(2)$. Note that in the usual parametrisation, the scalar matrix is 
\be
\mathcal{M}_{ij} = e^\Phi \begin{pmatrix} 
(C_0)^2 + e^{-2\Phi} &  C_0 \\ C_0 & 1 \end{pmatrix} \,,
\ee
where $\Phi$ is the string dilaton and $C_0$ the R-R zero form. Picking a vielbein 
\be
h^{\bi}{}_i = e^{\Phi/2} \begin{pmatrix} e^{-\Phi} & 0 \\ C_0 & 1 \end{pmatrix} \,,
\ee
one finds that the components of $\Gamma_{ab}{}_i{}^{j}$ are 
\be
\Gamma_{ab}{}_i{}^{j} = \begin{pmatrix} - \frac{1}{2} \partial_{ab} \Phi &  C_0 \partial_{ab} \Phi + \partial_{ab} C_0 \\ 0 & \frac{1}{2} \partial_{ab} \Phi \end{pmatrix} 
\,.
\label{eq:IIBScalarFluxExpl}
\ee
In general one may wish to introduce a dual field in place of $C_0$.

\subsection{Decomposition of the torsion irreps}

We can now, as before, express the torsion irreps in terms of these spacetime tensors (again, see
appendix \ref{seDetails} for the intermediate stages of the calculation). Note that the covariant
derivatives appearing in these expressions include a contribution from the scalar flux, so for
instance
\be
\tilde \nabla_{\mu i} V^j_\nu \equiv \tilde\partial_{\mu i} V^j_\nu + \tilde\Gamma_{\mu i
\nu}{}^\rho V_\rho^i + \tilde\Gamma_{\mu i k}{}^j V^k_\mu \,.
\ee

\vspace{1em}
\noindent\textbf{$\mathbf{15}$:}
We have
\begin{equation}
 \begin{split}
  S_{\mu\nu} &=  4 e \tilde\nabla_{k(\mu} V_{\nu)}{}^{k} - 2e \epsilon_{\kappa \lambda (\mu} T^{\kappa \lambda}{}_{\nu)} \,, \\
  S_{\mu i} & = - 2 \tilde \Gamma_{k\mu i}{}^k + 2 \mathcal{Q}^\rho{}_{i\mu \rho} + 2 \hat\nabla_{ki} V_\mu^k  + e^{-1} W_i^\nu S_{\mu \nu} \,, \\
  S_{ij} &= - 4 R_{ij} - 4 \hat \Gamma_{k(ij)}{}^{k} + 2 e^{-1} W_{(i}^\mu S_{j)\mu} -e^{-2} W_i^\mu W_j^\nu S_{\mu \nu} \,.
 \end{split}
\label{eq:IIBCurvedS}
\end{equation}
This gives the decomposition into $(\mathbf{6} , \mathbf{1} ) \oplus (\mathbf{3} , \mathbf{2} ) \oplus (\mathbf{1},\mathbf{3})$ of $\Sl(3) \times \Sl(2)$.

\vspace{1em}
\noindent\textbf{$\mathbf{10}$:}
We have
\begin{equation}
 \begin{split}
  \tau_{\mu\nu} &= \frac{1}{2} e \epsilon_{\kappa\lambda[\mu} {T}^{\kappa\lambda}{}_{\nu]} - \tilde\nabla_{i[\mu} V_{\nu]}{}^{i} - \frac{3}{2} \epsilon_{\mu \nu \rho} \partial^{\rho} \phi \,, \\
  \tau_{\mu i} &= -\frac{1}{2} \hat\nabla_{ij} V_{\mu}{}^{j} - \frac{1}{2} \mathcal{Q}^{\rho}{}_{i\mu\rho} - \frac{1}{2} \tilde \Gamma_{\mu ji}{}^{j} + \nT_{i\mu}  - \frac{3}{2} \tpartial_{\mu i} \phi + e^{-1} W_i^\nu \tau_{\mu \nu}  \,, \\
  \tau_{ij} &= - \frac{1}{2} e^{-1} R^{\rho}{}_{\rho ij} - \frac{3}{2} e^{-1} \hat\partial_{i j} \phi - 2 e^{-1} W_{[i}^\nu \tau_{j]\nu} - e^{-2} W_i^\mu W_j^\nu \tau_{ij} \,.
 \end{split}
\end{equation}
Here we have terms in the $(\mathbf{\bar{3}} , \mathbf{1} ) \oplus (\mathbf{3} , \mathbf{2} ) \oplus (\mathbf{1},\mathbf{1})$ of $\Sl(3) \times \Sl(2)$.

\vspace{1em}
\noindent\textbf{$\mathbf{40}$:}
We have 
\be
\begin{split} 
\widetilde{T}_{\mu \nu \rho}{}^i & = e^2 \epsilon_{\mu \nu \rho} \nabla^\lambda V_{\lambda}^i \\
& = e^2 \epsilon_{\mu \nu \rho} \left( \frac{1}{3!} \epsilon_{\kappa \lambda \sigma} H^{i \kappa \lambda
\sigma} - \frac{1}{2} \epsilon_{\kappa \lambda \sigma} B^{i \kappa \lambda} T^{\sigma \tau}{}_{\tau}
+ \frac{1}{2} \epsilon_{\kappa \lambda \sigma} \Gamma^{\sigma}{}_j{}^i B^{j\kappa \lambda} 
\right) \,,
\end{split}
\label{eq:IIBCurvedT1} 
\ee
\be
\widetilde{T}_{\mu \nu \rho}{}^{\lambda} = e \epsilon_{\mu \nu \rho} 
\left( \frac{2}{3} T^{\lambda \kappa}{}_{\kappa} 
- \frac{1}{3} \epsilon^{\lambda \kappa \sigma} \tilde\nabla_{i \kappa} V_\sigma^i 
\right) 
- e^{-1} W_i^\lambda \widetilde{T}_{\mu \nu \rho}{}^i \,,
\label{eq:IIBCurvedT2} 
\ee
\be
\begin{split}
\widetilde{T}_{\mu \nu i}{}^j & = 2 e \tilde \nabla_{i[\mu} V_{\nu]}^j - e \epsilon_{\mu \nu \lambda} \Gamma^\lambda{}_i{}^j + \frac{2}{3} e \delta^j_i \left( \frac{1}{2} \epsilon_{\kappa\lambda[\mu} T^{\kappa\lambda}{}_{\nu]} - \tilde \nabla_{k[\mu} V_{\nu]}^k \right)\\
& \qquad + e^{-1} W_i^\rho \widetilde{T}_{\mu \nu \rho}{}^j \,,
\end{split}
\label{eq:IIBCurvedT3} 
\ee
\be
\begin{split}
\widetilde{T}_{\mu \nu i}{}^\rho & = \mathcal{Q}^\rho{}_{i \mu \nu} - \frac{2}{3} \delta^\rho_{[\mu} \left( \nT_{|i|\nu]} + \hat\nabla_{|ij|} V_{\nu]}^j - 2 \mathcal{Q}^\lambda{}_{|i| \nu]\lambda} + \tilde \Gamma_{\nu]ji}{}^j \right) \\
 &  \qquad + e^{-1} W_i^\lambda \widetilde{T}_{\mu \nu \lambda}{}^\rho - e^{-1} W_j^\rho \widetilde{T}_{\mu \nu i}{}^j + e^{-2} W_j^\rho W_i^\lambda \widetilde{T}_{\mu \nu \lambda}{}^j \,,
\end{split} 
\label{eq:IIBCurvedT4} 
\ee
\be
\begin{split}
\widetilde{T}_{\mu i j}{}^k & =  \hat\nabla_{ij} V_\mu^k - 2 \tilde \Gamma_{\mu [ij]}{}^k + \frac{2}{3} \delta^k_{[i} \left( \hat\nabla_{j]l} V_\mu^l + \mathcal{Q}^\rho{}_{j]\mu \rho} + \tilde \Gamma_{|\mu|l|j]}{}^l - 2 \nT_{j]\mu} \right) \\
 &  \qquad + 2 e^{-1} W_{[i}^\lambda \widetilde{T}_{j]\mu \lambda}{}^k - e^{-2} W_i^\kappa W_j^\lambda \widetilde{T}_{\mu \kappa \lambda}{}^k \,,
\end{split} 
\label{eq:IIBCurvedT5} 
\ee
\be
\begin{split}
\widetilde{T}_{\mu i j}{}^\nu & = R^\nu{}_{\mu i j} - \frac{1}{3} \delta^\nu_\mu R^\rho{}_{\rho i j} \\
 & \qquad - e^{-1} W_k^\nu \widetilde{T}_{\mu i j}{}^k + 2 e^{-1} W_{[i}^\lambda \widetilde{T}_{j]\mu \lambda}{}^\nu - e^{-2} W_i^\kappa W_j^\lambda \widetilde{T}_{\mu \kappa \lambda}{}^\nu \\ & \qquad
+ 2 e^{-2} W_k^\nu W_{[i}^\lambda \widetilde{T}_{j]\mu\lambda}{}^k - e^{-3} W_i^\kappa W_j^\lambda W_k^\nu \widetilde{T}_{\mu \kappa \lambda}{}^k \,.
\end{split} 
\label{eq:IIBCurvedT6} 
\ee

The irreducible representations are 
\be
\begin{split}
 (\mathbf{8} , \mathbf{1} ) : & \quad R^\nu{}_{\mu ij} - \mathrm{trace}\,, \\ 
 (\mathbf{\overline{6}} , \mathbf{2} ) : & \quad 
\mathcal{Q}^\rho{}_{i \mu \nu} - \mathrm{trace} 
\,, \\ 
 (\mathbf{\overline{3}} , \mathbf{3} ) : & \quad 
2 \tilde \nabla_{i[\mu} V_{\nu]}^{j} - \epsilon_{\mu \nu \rho} \Gamma^{\rho}{}_i{}^k  -
\mathrm{trace} 
 \,, \\ 
 (\mathbf{3} , \mathbf{2} ): & \quad \hat{\nabla}_{ik} V^k_{\mu} - 2 \tilde\Gamma_{\mu[ij]}{}^k \,, \\ 
  (\mathbf{\overline{3}} , \mathbf{1} ) : & \quad T^{\mu \nu}{}_\nu \,, \\ 
(\mathbf{1},\mathbf{2} ) : &  \quad H^{i \mu \nu \rho} \,. \\ 
\end{split} 
\ee

\section{IIB flux examples}

In this final section of the paper, we will present some straightforward examples of duality chains connecting geometric and non-geometric fluxes in type IIB. 

\subsection{Duality chains and toy model} 

To generate duality chains in the IIB parametrisation, we again introduce an $\SLf$ duality element, which implements a pair of Buscher transformations (plus an interchange of the dualised coordinates).

As before, let $n_\mu$ point along the direction not being dualised and introduce $\bar{n}^\mu$ with $ n_\mu \bar{n}^\mu = 1$. We also need a two-component vector $m^i$, which should be taken to point along the $i = 1$ direction for a normal Buscher T-duality and along the $i = 2$ direction for its S-dual. Introduce $\bar{m}_i$ such that $m^i \bar{m}_i =  1$. Then we can take
\begin{equation}
 U^a{}_b = \begin{pmatrix}
  \delta^\mu{}_\nu - \bar{n}^\mu n_\nu & \bar{n}^\mu m_j \\
  -m^i n_\nu & \delta^i{}_j - m^i m_j
 \end{pmatrix} \,.
\label{eq:IIBBuscher}
\end{equation}
If we label our coordinates $x,y,z$ as before, and take $n_z=1$, $m^1=1$, then the effect of this duality is to exchange a $z$ index for a S-duality $1$ index, and a $1$ index for a $z$, up to a sign: letting $\alpha = x,y$ we would have $\tilde V^\alpha = V^\alpha, \tilde V^z = V^1, \tilde V^1 = - V^z$ and $\tilde V^2 = V^2$ for $\tilde V^a = U^a{}_b V^b$. Similarly, the effect on a lower index is to give $\tilde V_z = V_1$, $\tilde V_1 = V_z$ and the rest unchanged. 

As well as Buscher type transformations, we can also generate new fluxes using S-duality. These transformations are embedded in $\SLf$ in the obvious way:
\be
U^a{}_b = \begin{pmatrix} 
\delta^\mu{}_\nu & 0 \\
0 & A^i{}_j 
\end{pmatrix} \, , \quad \quad  A^i{}_j \in \mathrm{SL}(2) \,.
\ee
The basic $S$-duality inversion is generated by 
\be
A^i{}_j = \begin{pmatrix} 0 & - 1  \\ 1 & 0 \end{pmatrix} \,,
\ee
and we will denote the corresponding $\SLf$ transformation by $S$. We can immediately see for instance that our three-form fluxes $H^{i\mu \nu \rho}$ form a natural doublet under S-duality, as do their Buscher
duals, the Q-fluxes
$\mathcal{Q}^\mu{}_{i\nu \rho}$. Similarly, the symmetric R-fluxes $R_{ij}$ (which are Buscher dual to geometric flux) mix under S-duality
transformations.

The NS-NS sector duality chain of section \ref{se:NSNStoy} is of course available to us in the IIB theory, with the obvious difference that we are only allowed to do two Buscher dualities at a time. Just as in the M-theory case, the chain thus splits between the two irreducibles. 

Consider first the irreducible $\widetilde{T}_{abc}{}^d$, whose decomposition into IIB fluxes is given in equations \eqref{eq:IIBCurvedT1} - \eqref{eq:IIBCurvedT6}. Let us consider the toy set-up with coordinates $x,y,z$. Note that the coordinates $x_\mu$ in the IIB extended theory can be exchanged under duality for winding coordinates associated either to the NS-NS sector, $\tilde x^{\mu 1}$, or the Ramond-Ramond sector, $\tilde x^{\mu 2}$. We will denote the T-duality elements that do this for $x,y$ and their duals by $T_{xy,1}$ and $T_{xy,2}$, respectively. The effect of these elements on a U-duality tensor is, as noted above, to exchange the $z$ index with the $1$ or $2$ index, respectively. Note that $T_{xy,2} = S^{-1} T_{xy,1} S$. 

Let us start with a configuration with three-form NS-NS flux, $H^{1 xyz}$. This corresponds to the $\widetilde{T}_{xyz}{}^1$ component of the irreducible. Acting with T-duality on $x,y$ leads of course to $Q$-flux, corresponding to a non-zero $\widetilde{T}_{xy 1}{}^z$. Acting with S-duality gives the same picture, but in terms of Ramond-Ramond flux leading to a Ramond-Ramond $Q$-flux, which in our notation is $\mathcal{Q}^z{}_{2xy}$ (in the literature this has been referred to as $P^z{}_{xy}$ \cite{Aldazabal:2010ef}).

We can further act with $T_{xz,2}$ on the $\mathcal{Q}^z{}_{1xy}$ configuration or with $T_{xz,1}$ on the $\mathcal{Q}^z{}_{2xy}$ one, to reach the $\widetilde{T}_{x 1 2}{}^z$ component, which corresponds to a configuration with the non-vanishing $R$-flux, $R^z{}_{x12} \neq 0$. This is not the usual $R$-flux, but the novel type defined in \eqref{eq:IIBRNew}. This involves a duality acting on the non-isometry $z$ direction, and so indeed is expected to give a non-locally geometric flux. 

Alternatively, one can generate scalar flux \eqref{eq:IIBScalarFlux} by acting with $T_{xy,2}$ on the $\mathcal{Q}^z{}_{1xy}$ configuration, which leads to the $\widetilde{T}_{xy 1}{}^2$ component containing a non-vanishing $\Gamma^z{}_1{}^2$. 

Let us show how the latter two examples work in practice, in the context of the toy model. We start with the non-geometric NS-NS $Q$-flux solution in non-geometric frame:
\be
\begin{split}
ds^2 & = dx^2 + dy^2 + dz^2 \,, \\
\beta^2 & = N z \partial_x \wedge \partial_y \,.
\end{split} 
\label{eq:NSNSQsoln}
\ee
Let us act first with the $T_{xy,2}$ transformation. This produces a configuration in which the spacetime metric is unchanged and there are no two-form or bivectors present. However, there is a non-trivial matrix of scalars, giving 
\be
\begin{split}
d\tilde s^2 & = dx^2 + dy^2 + dz^2 \,, \\
C_0 & = - N z \,, \\
\Phi & = 0  \,.
\end{split} 
\ee
Here $\Phi$ is the string dilaton and $C_0$ is the Ramond-Ramond zero form. We see that the latter has a constant one-form flux, $F_1 = -N$. This corresponds to a non-zero scalar flux $\Gamma^{z}{}_1{}^2$, as can be seen by checking the explicit decomposition \eqref{eq:IIBScalarFluxExpl}.

Now, act on \eqref{eq:NSNSQsoln} with $T_{xz,2}$. This does not change the form of the solution, but changes which coordinates we are viewing as physical:
\be
\begin{split}
ds^2 & = d\tilde x^2 + d y^2 + d \tilde z^2 \,, \\
\beta^2 & = N z \partial_{\tilde x} \wedge \partial_{ y} \,.
\end{split} 
\label{eq:IIBNewRsoln}
\ee
We see that we are in the by now familiar situation of having an explicit dependence on what is now
a dual coordinate, $z$. This has a similar form to that of the usual $R$-flux background in the
NS-NS sector, \eqref{eq:NSNSRsoln}, however the coordinates $\tilde x,\tilde y$ appearing here are
not the usual dual coordinates (but rather their S-duals). To avoid becoming confused about which coordinates are which, rewrite the above as 
\be
\begin{split}
ds^2 & = d x^2 + d y^2 + d  z^2 \,, \\
\beta^2 & = - N \tilde{x}^{x2} \partial_{x} \wedge \partial_{ y} \,,
\end{split} 
\label{eq:IIBNewRsoln2}
\ee
where we have noted that the original coordinate $z$ becomes after the Buscher transformation the $\tilde x^{2x}$ coordinate from the point of view of this frame. This makes it easy to see that we have $\tilde \partial_{x 2} \beta_{xy} = -N$. Referring to the definitions of the two types of $R$-flux tensors, \eqref{eq:IIBROld} and \eqref{eq:IIBRNew}, we see that the former vanishes, and we have 
\be
R^z{}_{x12} = - N \,,
\ee
as expected from the duality chain. 

\begin{figure}[h]
\centering
\begin{tikzpicture}

\node [left] at  (-2.5,0) { $H^{1xyz}$} ;
\draw[->] (-1,0) -- (-0.6,0);
\draw[->]  (-1,0) -- (-2.4,0);
\node at (0,0) {$\mathcal{Q}^z{}_{1 xy}$};

\draw[->] (-3,-1) -- (-3,-0.3);
\draw[->] (-3,-1) -- (-3,-2);
\node [right] at (-3,-1.25) {$S$};
\node [below] at (-3,-2) {$H^{2xyz}$};
\node at (-1.5,-2) {$T_{xz,2}$};

\draw[->] (-1,-2.3) -- (-0.6,-2.3);
\draw[->]  (-1,-2.3) -- (-2.4,-2.3);
\node [above] at (-1.5,0) {$T_{xy,1}$};

\draw[->] (1,0) -- (0.6,0);
\draw[->] (1,0) -- (2.4,0);
\node [above] at (1.5,0) {$T_{xy,2}$};

\draw[->] (0,-1) -- (0,-0.3);
\draw[->] (0,-1) -- (0,-2);
\node [right] at (0,-1.25) {$S$};

\draw[->] (2.95,-1) -- (2.95,-0.3);
\draw[->] (2.95,-1) -- (2.95,-2);
\node [right] at (2.95,-1.25) {$S$};

\node [below] at (0,-2) {$\mathcal{Q}^z{}_{2 xy}$};

\draw[->] (1,-2.3) -- (0.6,-2.3);
\draw[->] (1,-2.3) -- (2.4,-2.3);
\node at (1.5,-2) {$T_{xz,1}$};

\node [below] at (2.95,-2) {$R^z{}_{x21}$};
\node [] at (2.95,0) {$R^z{}_{x12}$};
\end{tikzpicture}
\caption{Duality relations involving $3$-form, $Q$- and new $R$-flux in IIB.}
\label{fig:IIBToyduality}
\end{figure}

It is clear that the duality chains can be made more intricate, and that there are multiple paths between different backgrounds. For instance, we could also have obtained the $R$-form flux starting from a Ramond-Ramond scalar flux via:
\begin{equation}
 \Gamma^z{}_{1}{}^2 \, \xleftrightarrow{\,\,\,S\,\,\,} \, \Gamma^z{}_{2}{}^1 \, \xleftrightarrow{T_{yz,1}} \, R^x{}_{y12} \,. 
\end{equation}
Finally, let's consider the other irreducible, $S_{ab}$, given in terms of fluxes in \eqref{eq:IIBCurvedS}. A configuration with non-zero geometric flux $T^{yz}{}_x$ will have non-zero $S_{xx}$ component. The Buscher transformation $T_{yz,1}$ involving the non-isometry direction $z$ will then lead to a non-zero $S_{11}$ component, which means that we will have the usual non-geometric $R$-flux, $R_{11}$, as defined in \eqref{eq:IIBROld} (from which it is immediately clear that this component is the usual $R$-flux). Acting with the basic $S$-duality element then gives a non-zero $R_{22}$, which is just the $R$-flux defined for the Ramond-Ramond sector. 

\subsection{The $5^2_2$ solution and its S-dual} 

We can again illustrate a realistic example of how this works. This time we make use of the $5_2^2$
brane, which may be obtained by T-duality from the NS5 brane. As such it exists in both IIA and IIB
supergravity: the IIA form of the solution can in fact be obtained by reduction of the $5^3$
solution in M-theory. Hence the analysis of this brane is very similar to what we did before. Let us
compactify two transverse directions of the NS5. Carrying out a Buscher duality along one of these
directions gives the Kaluza-Klein monopole, and then carrying out an additional Buscher duality along the other direction gives the $5_2^2$. The solution has been comprehensively analysed in \cite{deBoer:2010ud, deBoer:2012ma}, and can be written as 
\be
\begin{split} 
ds^2&  = H \left( dr^2 + r^2 d\theta^2 \right) + HK^{-1} \left( dx^2 + dy^2 \right) + d\vec{x}_6^2 
\,,
\\
B_2 &= - \theta \sigma K^{-1} dx \wedge dy 
\,, 
\\  e^{\Phi} & = H^{1/2} K^{-1/2} \,,
\end{split}
\label{eq:522} 
\ee
where the function $H$ results from taking the original harmonic function of the NS5 and smearing on the compact directions, $\tilde x$ and $\tilde y$, which are T-dual to the compact directions $x$ and $y$:
\be
H = h_0 + \sigma \log \frac{\mu}{r} \,,
\ee
and $\mu$ is some cut-off and $h_0$ a bare quantity \cite{deBoer:2012ma}. The non-geometric properties of the background are due to the function $K$, which depends explicitly on the circular coordinate $\theta$,
\be
K = H^2 + \sigma^2 \theta^2 \,.
\ee
For $\theta \sim \theta + 2 \pi$ we have to act with a duality transformation that corresponds to a shift of a $\beta$ field. This cannot be realised on the above fields in terms of diffeomorphisms and $B$-field gauge transformations. If we change frame, replacing the two-form with a bivector, then we obtain a solution that looks geometric \cite{Hassler:2013wsa,Geissbuhler:2013uka}, 
\be
\begin{split}
ds^2&  = H \left( dr^2 + r^2 d\theta^2 \right) + H^{-1} \left( dx^2 + dy^2 \right) + d\vec{x}_6^2
\,,
\\ 
\beta^2 & = 
 \theta \sigma \partial_x   \wedge \partial_y \,, \\
e^{\Phi} & = H^{-1/2} \,.
\end{split}
\label{eq:522nongeo}
\ee
Due to the bivector with $\beta_{xy} = \theta\sigma$ this solution is thought of as carrying
non-geometric $Q$-flux.

We will now see that, as expected, the same holds after an S-duality.  

If we start from the IIB NS5 brane then the $5_2^2$ brane will also exist in IIB, with no R-R fields
turned on. If we act with the simple S-duality
\be
S = \begin{pmatrix} 0 & -1 \\ 1 & 0 \end{pmatrix} \,,
\ee 
then the net effect will be to exchange the $B_2$ field for a $C_2$ field. The resulting solution is
known as the $5_3^2$ \cite{deBoer:2012ma}, and can be written (in Einstein frame, note that \eqref{eq:522} is given in string frame) as
\be
\begin{split} 
ds^2_E&  = H^{3/4} K^{1/4} \left( dr^2 + r^2 d\theta^2 \right) + H^{3/4} K^{-3/4} \left( dx^2 + dy^2 \right) + H^{-1/4} K^{1/4} d\vec{x}_6^2 
\,,
\\
B_{2}^{i} & =
\begin{pmatrix} 
B_2 \\ 
C_2 
\end{pmatrix} = 
\begin{pmatrix} 
0 \\
- \theta \sigma K^{-1} dx \wedge dy 
\end{pmatrix} 
\,, 
\\  e^{\Phi} & = H^{-1/2} K^{1/2} \,.
\end{split}
\label{eq:S522} 
\ee
We can describe this in terms of the $\SLf$ exceptional field theory by supplementing the $\theta,x,y$ directions with seven dual coordinates. We have a choice of two parametrisations of the generalised vielbein \eqref{eq:IIBGenViel}, and hence the generalised metric, either using $V_\mu^i$ with $W^\mu_i =0$, or using $W^\mu_i$ with $V_\mu^i = 0$. By evaluating the generalised metric in the different parametrisations as in \eqref{eq:IIBChangeOfFrame}, we can straightforwardly read off the definitions of the various fields in the dual frame.

\begin{figure}[h]
\centering
\begin{tikzpicture}

\node [left] at  (-2.5,0) { NS5} ;
\draw[thick,->] (-1,0) -- (-0.5,0);
\draw[thick,->]  (-1,0) -- (-2.5,0);
\node [above] at (-1.5,0) {$T_{xy}$};
\node at (0,0) {$5_2^2$};

\draw[thick,->] (1,0) -- (0.5,0);
\draw[thick,->] (1,0) -- (2.5,0);
\node [above] at (1.5,0) {$S$};
\node [] at (2.95,0) {$5_3^2$};

\draw[->] (0,-1) -- (0,-0.5);
\draw[->] (0,-1) -- (0,-2);
\node [right] at (0,-1.25) {$ B \leftrightarrow \beta$};

\draw[->] (2.95,-1) -- (2.95,-0.5);
\draw[->] (2.95,-1) -- (2.95,-2);
\node [right] at (2.95,-1.25) {$ C \leftrightarrow \gamma$};

\node [below] at (0,-2) {$5_2^2$ };
\node [below] at (0,-2.5) {\small non-geo frame};

\draw[thick,->] (1,-2.3) -- (0.5,-2.3);
\draw[thick,->] (1,-2.3) -- (2.5,-2.3);
\node at (1.5,-2) {$S$};

\node [below] at (2.95,-2) {$5_3^2$};
\node [below] at (2.95,-2.5) {\small non-geo frame};
\end{tikzpicture}
\caption{Duality relations leading to a solution with non-geometric $Q$-flux in IIB.}
\label{fig:522Sduality}
\end{figure}
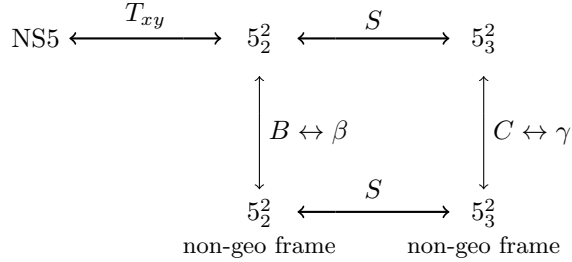

The dual frame form of the solution is
\be
\begin{split} 
d\tilde s^2_E&  = H^{5/4} \left( dr^2 + r^2 d\theta^2 \right) + H^{-3/4} \left( dx^2 + dy^2 \right) + H^{1/4}  d\vec{x}_6^2 
\,,
\\
\beta^2_i &=
\begin{pmatrix} 
\beta^2 \\ 
\gamma^2 
\end{pmatrix} = 
\begin{pmatrix}
 0 \\ - \theta \sigma \partial_x  \wedge \partial_y
\end{pmatrix} 
\,, 
\\  e^{\Phi} & = H^{1/2} \,.
\end{split}
\label{eq:S522dual} 
\ee
It is clear this solution carries a non-geometric flux associated to the derivative $\partial_\sigma \gamma^{xy}$ of the dual field $\gamma$, which we use in place of the usual R-R 2-form. This is just the S-dual of the usual non-geometric flux associated to the $5_2^2$ solution. One can check that the solution \eqref{eq:S522dual} is indeed related by S-duality to the $5_2^2$ solution in non-geometric frame, \eqref{eq:522nongeo}. The duality chain is summarised in figure \ref{fig:522Sduality}.

\subsection{IIB solution with $R$-flux} 

It is possible to obtain solutions with $R$-fluxes by various duality chains, all of which will at
some point need to include a duality along a non-isometry direction. For instance, one could act
with T-duality along the non-isometry direction of the $5_2^2$ solution to obtain the novel
$R$-flux, $R^\mu{}_{\nu 12}$, similar to the toy example discussed above. Alternatively, one could
start with the D7, which has a particularly simple S-duality monodromy affecting only the
Ramond-Ramond zero form and thus has scalar flux. Applying first an S-duality one obtains an
$S$-fold: further applications of T-duality lead to a background carrying the new $R$-flux,
$R^\mu{}_{\nu 12}$. Finally, a solution carrying the usual $R$-flux, $R_{ij}$, could be found
starting from a configuration with geometric flux, for instance the Kaluza-Klein monopole. 

\section{Conclusion}

In this paper, we have studied a geometric formalism for exceptional field theory which naturally contains information about all geometric and non-geometric fluxes. This geometric formalism made use of the generalised torsion of the Weitzenb\"ock connection: this generalised torsion can be used to naturally construct the action (by requiring invariance under local generalised Lorentz transformation), and unifies geometric and non-geometric fluxes into a single U-duality covariant object. As exceptional field theory reduces to both M-theory and type IIB, we obtain a unifying formalism for treating the fluxes of both these theories. 

We focused for simplicity on the U-duality group $\SLf$ and found new locally non-geometric fluxes which mix the R-R and NS-NS sector. We also showed how the new fluxes can be constructed by dualising geometric backgrounds. It would certainly be interesting to generalise the analysis here to the higher U-duality groups, leading to more complicated duality chains with more non-geometric fields.

It would be of interest to try and use the formalism developed here as a tool in generating backgrounds which \emph{cannot} be linked by duality to a known geometric solution. Such a background would be considered ``truly non-geometric''. In order to do so, it will be necessary to understand the consistency constraints, or equivalently, Bianchi identities, that the fluxes must obey. We leave it to a future work to present a full analysis of these constraints in terms of the spacetime fluxes we have identified.

Our formalism would also allow us to construct actions involving non-geometric fluxes. The non-geometric branes considered here would then be solutions of these actions. The actions would allow one to further study configurations involving dual fields, for instance, and would be useful for determining the effective potentials resulting from a Scherk-Schwarz reduction. It would be interesting to understand the phenomenological consequences of the new fluxes considered here.
Furthermore, the results presented here will help us understand how the non-commutativity /
non-associativity of strings and exotic branes
\cite{Lust:2010iy,Blumenhagen:2010hj, Blumenhagen:2011ph,Condeescu:2012sp, Mylonas:2012pg,
Andriot:2012vb,Bakas:2013jwa,
Davidovic:2013rma,
Hassler:2013wsa,Blair:2014kla
} 
in non-geometric
backgrounds generalise to M-theory or are modified in the presence of Ramond-Ramond fields. In
particular, it is interesting to note that the locally non-geometric flux in M-theory is not totally
antisymmetric, in contrast to the NS-NS case. This makes non-associative behaviour unlikely.
However, it may signal that a higher bracket structure, such as a Nambu bracket, is needed in the analysis.

We have seen in this paper that exceptional field theory provides a natural setting for studying non-geometric backgrounds. It would be interesting to study the generalised coordinate patching \cite{ Berman:2014jba, Papadopoulos:2014mxa} of the extended space necessary to fully define such backgrounds, as has been studied in the T-duality case \cite{Hohm:2012gk,Hohm:2013bwa,Park:2013mpa}.

\section*{Acknowledgements}

We would like to thank David Berman, Ralph Blumenhagen, Dieter L\"ust and Malcolm Perry for useful discussions. CB thanks St John's College, Cambridge, for their support. EM is funded by the National Research Foundation (NRF) of South Africa under grant CSUR13091742207. EM would also like to thank the Max Planck Institute Munich (Werner-Heisenberg Institute) for hospitality while part of this work was undertaken.

\appendix

\section{Generalised metrics and non-geometric frames}

\subsection{M-theory changes of frame}

The idea here is simple. The generalised metric itself is taken to be the fundamental field of the theory. The choice of physical fields is viewed as a choice of how to parametrise the generalised metric. This frees us from having to always use one particular set of fields, which in certain circumstances may be in fact unsuitable. 

The particular situation we are interested in will be changes of frame from a situation where, by acting with duality transformations, we have a description of a background in terms of the usual metric and the three-form, to a frame where we have an alternative metric and a dual trivector in place of the three-form.\footnote{The idea of parameterising the generalised metric of string theory in terms of a dual field was used in \cite{Grana:2008yw} and \cite{Andriot:2011uh} to study non-geometric backgrounds.}

The little metric that follows from the general form of the M-theory generalised vielbein, \eqref{eq:MTheoryE}, is
\be 
m_{ab} = e^{-\phi/2} 
\begin{pmatrix}
g^{-1/2} \left( g_{ij} + W_i V_j + V_i W_j + W_i W_j (1  + V^2) \right) & V_i + W_i (1+V^2) \\
V_j + W_j (1 + V^2) & g^{1/2} ( 1 + V^2 ) 
\end{pmatrix}\,.
\label{eq:LittleMetGeneral}
\ee
In the usual geometric description, we set $W_i =0$. In a non-geometric situation, we may have to take instead $V^i = 0$. The generalised metric remains the same in both expressions. The transformation from frame to frame can be realised as a generalised Lorentz transformation acting on the flat index of the generalised vielbein. There may be global issues in defining such a transformation. 

Using the expressions for the generalised metric in each frame, one can read off the definitions of the dual metric $\tilde{g}_{ij}$, trivector $\Omega^{ijk}$ and (via $e^{\phi} \equiv |g_7|^{1/7}$) the metric in the seven transverse directions, $\tilde{g}_{IJ}$, in terms of the original variables:
\be
\begin{split} 
\tilde{g}_{ij}  & = (1 + V^2)^{-1/3} \left( (1 + V^2 ) g_{ij} - V_i V_j \right)   \,, \\ 
\Omega^{ijk}  & = (1 + V^2)^{-1} g^{il} g^{jm} g^{kn} C_{kmn} \,, \\ 
\tilde{g}_{IJ}  & = (1+V^2)^{-1/3} g_{IJ}\,. 
\end{split}
\label{eq:MChangeOfFrame}
\ee

\subsection{IIB changes of frame}

The standard parametrisation involves a three-dimensional metric, $g^{\mu \nu}$, a pair of two-forms, $B^{i \mu \nu}$, two scalars packaged into a symmetric unit determinant two-by-two matrix, $\mathcal{M}_{ij}$, and the transverse metric $g^{IJ}$ (denoted with upper indices for consistency). Let us suppose we change frame to a parametrisation in which instead of two-forms we have a pair of bivectors $\beta_{i\mu \nu}$. Again, we denote the other quantities in the new frame with tildes. The general expression for the little metric, from the parametrisation \eqref{eq:IIBGenViel}, is
\be
m_{ab} = e^{-\phi/2}
\begin{pmatrix}
g^{1/2} ( g_{\mu \nu} + V_\mu^k V_{\nu k} ) & V_{\mu j} + W_{\mu j} + V_{\mu}^k V_{\rho k} W_{j}^\rho \\
V_{\nu i} + W_{\nu i} + V_{\nu}^k V_{\rho k} W_i^\rho & m_{ij}
\end{pmatrix} \,,
\ee
where
\begin{equation}
 m_{ij} = g^{-1/2} \left(
 \mathcal{M}_{ij} + W_i^\rho W_{j\rho} + V_{\rho i} W^{\rho}_j + V_{\rho j} W^{\rho}_i + W_i^\rho W_j^\sigma V_\rho^k V_{\sigma k} 
 \right) \,.
\end{equation}
To write down the expressions for the change from geometric to non-geometric frame, we first define the following determinants:
\be
|g_3| \equiv \det ( g^{\mu \nu} )
\quad , \quad
|g_7| \equiv \det ( g^{IJ} ) 
\quad , \quad
|g + V^2  | \equiv \det \left( g_{\mu \nu} + V_\mu^i \mathcal{M}_{ij} V_\nu^j \right) \,.
\ee
Here, 
\be
V^i_\mu \equiv \frac{1}{2} \epsilon_{\mu \nu \rho} B^{i \nu \rho} \,,
\ee
and we similarly would define
\be
W_i^\mu \equiv \frac{1}{2} \tilde\epsilon^{\mu \nu \rho} \beta_{i \nu \rho} \,,
\ee
for the dual field. 

We then have the following formulae for the quantities in the new frame: 
\be
\begin{split} 
\tilde g_{\mu \nu}  & = |g_3|^{-3/4} | g + V^2 |^{-3/4}\left( g_{\mu \nu} + V_\mu^i \mathcal{M}_{ij} V_\nu^j \right) \,,\\
\tilde g^{IJ} & = |g_3|^{-1/4} | g + V^2  |^{-1/4} g^{IJ} \,, \\
  \beta_{j\mu \nu}  & = |g_3|^{1/2} | g + V^2  |^{1/2} \tilde g_{\mu \rho} \tilde g_{\nu \sigma} \mathcal{M}_{jk} B^{k \rho\sigma}  \,, \\
 \tilde{\mathcal{M}}_{ij}& = |g_3|^{1/2} | g + V^2  |^{1/2} \mathcal{M}_{ij} - W_i^\mu W_j^\nu \tilde g_{\mu \nu}  \,.
\end{split} \label{eq:IIBChangeOfFrame}
\ee

\section{Details of the torsion decompositions}
\label{seDetails}

\subsection{M-theory}

We first give the components of the flattened Weitzenb\"ock connection, defined by
\be
\Omega_{\alpha \beta \gamma}{}^\delta \equiv E_\gamma{}^a \mathcal{D}_{\alpha \beta} E^\delta{}_a \,,
\ee
and evaluated using the parametrisation \eqref{eq:MTheoryE}. These involve
\be
D_{\alpha\beta} W_\mu = \mathcal{D}_{\alpha\beta} W_\mu + \Gamma_{\alpha\beta \mu}{}^\nu W_\nu -  \Gamma_{\alpha\beta \rho}{}^\rho W_\mu
= \frac{1}{3!} e_\mu{}^i \epsilon_{ijkl} \mathcal{D}_{\alpha\beta} \Omega^{jkl} \,,
\ee
\be
D_{\alpha\beta} V^\mu = \mathcal{D}_{\alpha\beta} V^\mu - \Gamma_{\alpha\beta \nu}{}^\mu V^\nu +  \Gamma_{\alpha\beta \rho}{}^\rho V^\mu
= \frac{1}{3!} e^\mu{}_i \epsilon^{ijkl} \mathcal{D}_{\alpha\beta} C_{jkl} \,,
\ee
using the flat partial derivatives \eqref{eq:FlatDer} and the spacetime Weitzenb\"ocks with flat indices, $\Gamma_{\alpha\beta \mu}{}^\nu \equiv e_\mu{}^i \mathcal{D}_{\alpha \beta} e^\nu{}_i$.
The generalised Weitzenb\"ock components are then
\be
\Omega_{\alpha\beta\mu}{}^\nu = \Gamma_{\alpha\beta\mu}{}^\nu - \frac{1}{2} \delta^\nu_\mu \Gamma_{\alpha\beta \lambda}{}^\lambda   + V^\nu D_{\alpha\beta} W_\mu
-\frac{1}{4} \delta^\nu_\mu \mathcal{D}_{\alpha \beta}\phi
\,,
\ee
\be
\Omega_{\alpha\beta\mu}{}^5 = D_{\alpha\beta} W_\mu \,,
\ee
\be
\Omega_{\alpha\beta5}{}^\mu = D_{\alpha\beta} V^\mu - V^\mu V^\nu D_{\alpha\beta} W_\nu \,,
\ee
\be
\Omega_{\alpha\beta5}{}^5 = - V^\mu D_{\alpha\beta} W_\mu + \frac{1}{2} \Gamma_{\alpha\beta \lambda}{}^\lambda 
-\frac{1}{4}  \mathcal{D}_{\alpha\beta}\phi
 \,.
\ee
We introduce the following notation to distinguish between the different types of derivatives. Derivatives flattened with the spacetime vielbein will be denoted with a bar:
\be
\bpartial_\mu \equiv e_\mu{}^i \partial_i \,, \qquad \bpartial_{\mu \nu} \equiv e_\mu{}^i e_\nu{}^j \partial_{ij} \,,
\ee
and objects (connections and torsions) built using these will also be barred. We have
\be
\mathcal{D}_\mu - V^\lambda \mathcal{D}_{\lambda \mu} = \bpartial_\mu \,.
\ee
Using this, we find for the $\mathbf{15}$, 
\be
S_{\mu \nu} = 4 \left( \Gamma_{\lambda(\mu \nu)}{}^\lambda - \bar{D}_{(\mu} W_{\nu)} \right) \,,
\ee
\be
S_{\mu 5} = 2 \left( T_{\lambda \mu}{}^\lambda + D_{\lambda \mu} V^\lambda + 2 V^\lambda \bar{D}_{(\mu} W_{\lambda)} \right)  \,,
\ee
\be
S_{55} = 4 \left( D_\mu V^\mu - V^\mu V^\nu D_{\mu} W_\nu\right) \,,
\ee
which leads to
\be
\begin{split} 
S_{\mu \nu} & = - 4 e^{\phi/2} \mathcal{Q}_{(\mu,\nu)} \,,
\\
S_{\mu 5} & = 2 e^{\phi/2} \left( \hat\partial_{\lambda \mu} V^\lambda +  {T}_{\lambda \mu}{}^\lambda \right) - V^\nu S_{\mu\nu}\,,
\\
S_{55} & = 4 e^{\phi/2} \partial_\mu V^\mu - 2 V^\mu S_{\mu 5} - V^\mu V^\nu S_{\mu \nu} \,.
\end{split} 
\ee
For the $\mathbf{10}$, 
\be
\tau_{\mu 5} = \frac{1}{2} T_{\lambda \mu}{}^\lambda - \frac{1}{2} D_{\lambda \mu} V^{\lambda} - V^\lambda \bar{D}_{[\mu} W_{\lambda]} \,,
\ee
\be
\tau_{\mu \nu} = - \Omega_{\lambda [\mu \nu]}{}^\lambda - \Gamma_{\mu \nu \lambda}{}^\lambda + \bar{D}_{[\mu} W_{\nu]} \,,
\ee
leading to
\be
\begin{split}
\tau_{\mu \nu} &  = e^{\phi/2}\left( \epsilon_{\mu \nu \kappa \lambda} \nT^{\kappa , \lambda} + \mathcal{Q}_{[\mu, \nu]}
- \frac{3}{2} \hat\partial_{\mu\nu} \phi \right)  
 \,,\\
\tau_{\mu 5} & = e^{\phi/2} \left( - \frac{1}{2} \hat\partial_{\lambda \mu} V^\lambda - \frac{1}{2}{T}_{\mu
\lambda}{}^\lambda - \frac{3}{2}  {\partial}_{\mu} \phi\right) - V^\nu \tau_{\mu \nu} \,.
\end{split}
\ee
For the $\mathbf{40}$, 
\be
\begin{split} 
\widetilde{T}_{\mu \nu \rho}{}^5 & = 3 D_{[\mu \nu} W_{\rho]} \,,
\end{split} 
\ee

\be
\begin{split} 
\widetilde{T}_{\mu \nu 5}{}^5 & = - 3  V^\lambda {D}_{[\mu \nu} W_{\lambda]} + \Gamma_{[\mu \nu \lambda]}{}^\lambda
- \frac{4}{3} \left( \bar{D}_{[\mu} W_{\nu]} + \Gamma_{\lambda[\mu \nu]}{}^{\lambda} \right) \,,
\end{split} 
\ee

\be
\begin{split} 
\widetilde{T}_{\mu \nu \rho}{}^\lambda & = 3  V^\lambda {D}_{[\mu \nu} W_{\rho]} + 3 \Gamma_{[\mu \nu \rho]}{}^\lambda 
+ 2 \bar{D}_{[\mu} W_\nu \delta_{\rho]}^\lambda - 2 \Gamma_{\kappa [ \mu \nu}{}^\kappa \delta_{\rho]}^\lambda
- 2 \delta^\lambda_{[\mu} \Gamma_{\nu \rho] \kappa}{}^\kappa \,,
\end{split} 
\ee

\be
\begin{split} 
\widetilde{T}_{\nu \rho 5}{}^\mu & = 
 D_{\nu\rho} V^\mu  - T_{\nu\rho}{}^\mu
- 3  V^\mu V^\kappa {D}_{[\nu\rho} W_{\kappa]}
 - 2 V^\mu \bar{D}_{[\nu} W_{\rho]} 
\\ &  
+ \frac{2}{3} \delta^\mu{}_{[\nu} D_{\rho]\kappa} V^\kappa 
 - \frac{2}{3} \delta^\mu{}_{[\nu} T_{\rho]\kappa}{}^\kappa
  - \frac{2}{3} \delta^\mu{}_{[\nu} V^\kappa \bar{D}_{\rho]} W_{\kappa}
  + \frac{2}{3} \delta^\mu{}_{[\nu} V^\kappa \bar{D}_{\kappa]} W_{\rho} \,,
\end{split}
\ee
leading to
\be
\begin{split} 
\widetilde{T}_{\mu \nu \rho}{}^5 & =-  e^{\phi/2} \epsilon_{\mu \nu \rho \kappa} L^\kappa \,,
\end{split} 
\ee
\be
\widetilde{T}_{\mu \nu 5}{}^5 = e^{\phi/2} \left( - \frac{4}{3}   \mathcal{Q}_{[\mu,\nu]} - \frac{1}{3}  \epsilon_{\mu \nu \kappa
\lambda} \nT^{\kappa,\lambda} 
\right)
- V^\lambda \widetilde{T}_{\mu \nu \lambda}{}^5 \,,
\ee
\be
\widetilde{T}_{\mu \nu \rho}{}^\lambda = 
e^{\phi/2} \left( 
2 \mathcal{Q}_{[\mu,\nu } \delta_{\rho]}^\lambda + \epsilon_{\mu \nu \rho \kappa}
\nT^{\kappa,\lambda} 
+ 2  \delta^\lambda_{[\mu} \epsilon_{\nu \rho] \sigma \kappa} \nT^{\sigma , \kappa} \right) 
 + V^\lambda \widetilde{T}_{\mu \nu \rho}{}^5 \,,
\ee
\be
\begin{split}
\widetilde{T}_{\nu \rho 5}{}^\mu & = 
e^{\phi/2} \left( 
 \hat\partial_{\nu \rho} V^\mu + \frac{2}{3} \delta^{\mu}_{[\nu} \hat\partial_{\rho] \lambda} V^\lambda
- {T}_{\nu \rho}{}^\mu - \frac{2}{3}  \delta^{\mu}_{[\nu} {T}_{\rho]\lambda}{}^\lambda \right)
\\ & \quad + V^\mu \widetilde{T}_{\nu \rho 5}{}^5
- V^\lambda \widetilde{T}_{\nu \rho \lambda}{}^\mu 
+ V^\mu V^\lambda \widetilde{T}_{\nu \rho \lambda}{}^5\,.
\end{split} 
\ee

\subsection{IIB}

First, let us give the components of the flat Weitzenb\"ock. We have spacetime connections
\be
\Gamma_{\alpha \beta}{}^{\bmu}{}_{\bnu}  \equiv e^{\bmu}{}_\mu \mathcal{D}_{\alpha \beta} e_{\bnu}{}^\mu \,,
\ee
and similarly for the scalar coset vielbein $h_{\bi}{}^i$
\be
\Gamma_{\alpha \beta \bj}{}^{\bi} \equiv h^{\bi}{}_i \mathcal{D}_{\alpha \beta} h_{\bj}{}^i \,.
\ee
The components involve the following combinations
\be
D_{\alpha \beta} W^{\bmu}_{\bk} \equiv 
\mathcal{D}_{\alpha \beta} W^{\bmu}_{\bk} - \Gamma_{\alpha \beta \bk}{}^{\bj} W^{\bmu}{}_{\bj}
 + \Gamma_{\alpha \beta}{}^{\bmu}{}_{\bnu} W^{\bnu}_{\bk} 
- \Gamma_{\alpha \beta}{}^{\brho}{}_{\brho}
W^{\bmu}_{\bk}
= \frac{1}{2} h_{\bk}{}^k e^{\bmu}{}_\mu \epsilon^{\mu \nu \rho} \mathcal{D}_{\alpha \beta} \beta_{k \nu \rho} \,,
\ee
\be
D_{\alpha \beta} V_{\bmu}^{\bk} \equiv
\mathcal{D}_{\alpha \beta} V_{\bmu}^{\bk} + \Gamma_{\alpha \beta \bj}{}^{\bk} V_{\bmu}^{\bj} 
- \Gamma_{\alpha \beta}{}^{\bnu}{}_{\bmu} V_{\bnu}^{\bk} + 
\Gamma_{\alpha \beta}{}^{\brho}{}_{\brho} 
V_{\bmu}^{\bk} 
= \frac{1}{2} h^{\bk}{}_k e_{\bmu}{}^\mu \epsilon_{\mu \nu \rho} \mathcal{D}_{\alpha \beta} B^{k \nu \rho}  \,.
\ee
Then we have
\be
\Omega_{\alpha \beta \bnu}{}^{\bmu} = 
- \Gamma_{\alpha \beta}{}^{\bmu}{}_{\bnu}
+ \frac{1}{2} \Gamma_{\alpha \beta}{}^{\brho}{}_{\brho} \delta^{\bmu}_{\bnu} 
- V_{\bnu}^{\bk} D_{\alpha \beta} W^{\bmu}_{\bk} 
- \frac{1}{4} \delta^{\bmu}_{\bnu} \mathcal{D}_{\alpha \beta} \phi
\,,
\ee
\be
\Omega_{\alpha \beta \bi}{}^{\bmu} = D_{\alpha \beta} W^{\bmu}_{\bi} \,,
\ee
\be
\Omega_{\alpha \beta \bmu}{}^{\bi} = D_{\alpha \beta} V_{\bmu}^{\bi} - V_{\bmu}^{\bk}
V_{\brho}^{\bi} D_{\alpha \beta} W^{\brho}_{\bk}  \,,
\ee
\be
\Omega_{\alpha \beta \bj}{}^{\bi} = - \Gamma_{\alpha \beta \bj}{}^{\bi} 
- \frac{1}{2} \Gamma_{\alpha \beta}{}^{ \brho}{}_{\brho} \delta^{\bi}_{\bj} 
+  V_{\brho}^{\bi} D_{\alpha \beta} W^{\brho}_{\bi}
- \frac{1}{4} \delta^{\bi}_{\bj} \mathcal{D}_{\alpha \beta} \phi
 \,.
\ee
Then for the $\mathbf{15}$, 
\be
S_{\bmu \bnu} = - 4 \Gamma_{\brho ( \bmu}{}^{\brho}{}_{\bnu ) }
+ 4V_{(\bmu}^{\bk}  D_{\bnu) \brho} W^{\brho}_{\bk} 
+ 4  D_{\bk (\bmu} V_{\bnu)}^{\bk}
+ 4  V^{\bj}_{\brho} V^{\bk}_{(\bmu} D_{\bnu) \bj} W^{\brho}_{\bk} \,,
\ee
\be
\begin{split}
\frac{1}{2} S_{\bmu \bi} & = 
- \Gamma_{\bk \bmu \bi}{}^{\bk}  - \Gamma_{\brho \bi}{}^{\brho}{}_{\bmu}
 + \Gamma_{\bmu \bi}{}^{\brho}{}_{\brho} +  D_{\brho \bmu} W^{\brho}_{\bi} 
+ D_{\bk \bi} V^{\bk}_{\bmu}  \\
 & \quad + V_{\brho}^{\bk} D_{\bk \bmu} W^{\brho}_{\bi} 
- V_{\bmu}^{\bk} D_{\brho \bi} W^{\brho}_{\bk}
- V^{\bj}_{\bmu} V^{\bk}_{\brho} D_{\bk \bi} W^{\brho}_{\bj}  \,,
\end{split}
\ee
\be
S_{\bi \bj} = 
 4D_{\brho ( \bi} W^{\brho}_{\bj)} 
-4 \Gamma_{\bk (\bi \bj)}{}^{\bk} 
+  4 V_{\brho}^{\bk} D_{\bk( \bi } W^{\brho}_{\bj)} \,.
\ee
These lead to 
\begin{equation}
 \begin{split}
  S_{\bi\bj} &= - 4 e^{\phi/2} \left(  R_{\bi\bj} + \tGamma_{\bk(\bi\bj)}{}^{\bk} \right) \,, \\
  S_{\bi\bmu}
  &= 2 e^{\phi/2} \left( - \tGamma_{\bk\bmu\bi}{}^{\bk} +  \mathcal{Q}^{\brho}{}_{\bi\bmu\brho} +  \hat\partial_{\bk\bi} V_{\bmu}{}^{\bk} \right) - V_{\bmu}{}^{\bj} S_{\bi\bj} \,, \\
  S_{\bmu\bnu} &=  4 e^{\phi/2} \left( \tpartial_{\bk(\bmu} V_{\bnu)}{}^{\bk} - \frac{1}{2}
\epsilon_{\blambda\brho(\bmu} T^{\blambda\brho}{}_{\bnu)}\right) -2 V_{(\bmu}{}^{\bi} S_{\bnu)\bi} -
V_{\bmu}{}^{\bi}  V_{\bnu}{}^{\bj} S_{\bi\bj} \,.
 \end{split}
\end{equation}
Next, for the $\mathbf{10}$, 
\be
\tau_{\bmu \bnu} = \Gamma_{\brho [ \bmu}{}^{\brho}{}_{ \bnu ]} 
+ V_{[\bmu}^{\bk} D_{\bnu]\brho} W^{\brho}_{\bk}
- D_{\bk[\bmu} V_{\bnu]}^{\bk} 
 + V_{\brho}^{\bk} V_{[\bmu}^{\bj}  D_{\bnu] \bk} W_{\bj}^{\brho} - \frac{3}{2} \mathcal{D}_{\bmu \bnu} \phi\,,
\ee
\be
\begin{split} 
\tau_{\bmu \bi} & = - \frac{1}{2} \Gamma_{\bmu \bi}{}^{ \brho}{}_{\brho} 
- \frac{1}{2} \Gamma_{\brho \bi}{}^{\brho}{}_{ \bmu} - \frac{1}{2} \Gamma_{\bmu \bk \bi}{}^{\bk} \\ 
& 
- \frac{1}{2} D_{\brho \bmu} W_{\bi}^{\brho} 
-\frac{1}{2} V_{\brho}^{\bk} D_{\bk \bmu} W_{\bi}^{\brho} 
- \frac{1}{2} V_{\bmu}^{\bk} D_{\brho \bi} W^{\brho}_{\bk} 
\\ 
 & + \frac{1}{2} D_{\bk \bi} V_{\bmu}^{\bk} 
- \frac{1}{2} V_{\bmu}^{\bj} V_{\brho}^{\bk} D_{\bk \bi} W_{\bj}^{\brho} 
 - \frac{3}{2} \mathcal{D}_{\bmu \bi} \phi\,,
\end{split}
\ee
\be
\tau_{\bi \bj} = -  \Gamma_{\bi \bj}{}^{\brho}{}_{ \brho}
- D_{\brho[\bi} W_{\bj]}^{\brho} - V_{\brho}^{\bk} D_{\bk[\bi} W_{\bj]}^{\brho} 
 - \frac{3}{2} \mathcal{D}_{\bi \bj} \phi \,,
\ee
which lead to 
\begin{equation}
 \begin{split}
  \tau_{\bi\bj} &= - \frac{1}{2} e^{\phi/2} \left(  R^{\brho}{}_{\brho \bi \bj} + 3 \hat\partial_{\bi \bj} \phi \right)  \,, \\ 
  \tau_{\bmu\bi} &= -\frac{1}{2} e^{\phi/2} \left( 
 \hat\partial_{\bi\bj} V_{\bmu}{}^{\bj} +  \mathcal{Q}^{\brho}{}_{\bi\bmu\brho} +  \tGamma_{\bmu\bj\bi}{}^{\bj} 
-2  \nT_{\bi\bmu} + 3  \tpartial_{\bmu \bi} \phi \right) 
 + V_{\bmu}{}^{\bj} \tau_{\bi\bj}\,, \\
  \tau_{\bmu\bnu} &= -\frac{1}{2} e^{\phi/2} \left(- \epsilon_{\bar{\kappa} \blambda [\bmu} {T}^{\bar{\kappa} \blambda }{}_{\bnu]} + 2 \tpartial_{\bi[\bmu} V_{\bnu]}{}^{\bi} 
+  3 \epsilon_{\bmu \bnu \brho} \partial^{\brho} \phi \right)
+ 2 V^{\bi}_{[\bmu} \tau_{\bnu]\bi}  - V_{[\bmu}{}^{\bi} V_{\bnu]}{}^{\bj} \tau_{\bi\bj}
  \,.
 \end{split}
\end{equation}
Finally, for the $\mathbf{40}$, 
let us first simplify the calculation by noting that 
\be
\begin{split}
\widetilde{T}_{bcd}{}^a & = 3 \Omega_{[bcd]}{}^a - \delta^a_{[b} \Omega_{cd]e}{}^e  - 2
\delta^a_{[b|} \Omega_{e|cd]}{}^e \\
& = 3 \Omega_{[bcd]}{}^a -3 \delta^a_{[b} \Omega_{cd]e}{}^e 
+ 2 \delta^a_{[b} \tau_{cd]}  \,.
\end{split} 
\ee
Thus the novel terms will be the first two. One obtains the expressions
\be
\widetilde T_{\bmu \bnu \brho}{}^{\bi} = 3  D_{ [ \bmu \bnu } V_{\brho]}^{\bi}
- 3 V_{\bar{\lambda}}^{\bi} V_{[\brho}^{\bk} D_{\bmu \bnu]} W^{\bar{\lambda}}_{\bk} \,,
\ee
\be
\begin{split} 
\widetilde T_{\bmu \bnu \brho}{}^{\bar{\lambda}} & = 
- 3 \Gamma_{[\bmu \bnu}{}^{\blambda}{}_{ \brho]}
+ 3  \delta^{\bar{\lambda}}_{[\bmu} \Gamma_{\bnu \brho]}{}^{ \bar{\kappa} }{}_{\bar{\kappa}} 
- 3 V^{\bk}_{[\brho} D_{\bmu \bnu]} W^{\bar{\lambda}}_{\bk} 
+ 2 \delta^{\bar{\lambda}}_{[\bmu}  \left( \tau_{\bnu \brho]} + \frac{3}{2} \mathcal{D}_{\bnu \rho]} \phi \right) \,,
\end{split} 
\ee
\be
\begin{split} 
\widetilde T_{\bmu \bi \bj}{}^{\bk} & = 
- 2 \Gamma_{ \bmu [ \bi \bj ]}{}^{\bk} 
+ 2 V_{\brho}^{\bk} D_{\bmu [ \bi} W_{\bj]}^{\brho} 
+ D_{\bi \bj} V_{\bmu}^{\bk} - V_{\brho}^{\bk} V_{\bmu}^{\bar{l}} D_{\bi \bj} W_{\bar{l}}^{\brho}
+ \frac{4}{3} \delta_{[\bi}^{\bk} \left( \tau_{\bj] \bmu} + \frac{3}{2} \mathcal{D}_{\bj] \bmu } \phi \right) \,,
\end{split} 
\ee
\be
\begin{split} 
\widetilde T_{\bmu \bnu \bi}{}^{\bj} = 
- \Gamma_{\bmu \bnu \bi}{}^{\bj} 
+ V_{\brho}^{\bj} D_{\bmu \bnu} W_{\bi}^{\brho} 
+ 2 D_{\bi [ \bmu} V_{\bnu]}^{\bj}
- 2 V_{\brho}^{\bj} V^{\bk}_{[\bmu} D_{\bnu]\bi} W_{\bk}^{\brho} 
+ \frac{2}{3} \delta_{\bi}^{\bj} \left( \tau_{\bmu \bnu} + \frac{3}{2} \mathcal{D}_{\bmu \bnu } \phi \right) \,,
\end{split} 
\ee
\be
\begin{split} 
\widetilde T_{\bmu \bnu \bi}{}^{\brho} & = 
- 2 \Gamma_{\bi[\bmu}{}^{\brho}{}_{ \bnu]}
+ 2 \delta^{\brho}_{[\bmu} \Gamma_{\bnu ] \bi}{}^{\blambda}{}_{\blambda} 
+ 
D_{\bmu \bnu} W_{\bi}^{\brho} 
- 2 V^{\bk}_{[\bmu} D_{\bnu] \bi} W_{\bk}^{\brho}
+ \frac{4}{3} \delta^{\brho}_{[\bmu} \left( \tau_{\bnu]\bi} + \frac{3}{2} \mathcal{D}_{\bnu]\bi } \phi \right) \,,
\end{split} 
\ee
\be
\begin{split} 
\widetilde T_{\bmu \bi \bj}{}^{\bnu} & = 
- \Gamma_{\bi \bj}{}^{\bnu}{}_{ \bmu}
+ \delta_{\bmu}^{\bnu} \Gamma_{\bi \bj}{}^{\brho}{}_{\brho}
+ 
2 D_{\bmu [ \bi} W_{\bj]}^{\bnu} 
- V_{\bmu}^{\bk} D_{\bi \bj} W_{\bk}^{\bnu}
+ \frac{2}{3} \delta^{\bnu}_{\bmu} \left( \tau_{\bi \bj} + \frac{3}{2} \mathcal{D}_{ \bi \bj} \phi \right)\,.
\end{split} 
\ee
These lead to the following expressions:
\be
\widetilde{T}_{\bmu \bi \bj}{}^{\bnu} = 
e^{\phi/2} \left( 
R^{\bnu}{}_{\bmu \bi \bj} 
- \frac{1}{3} \delta^{\bnu}_{\bmu}R^{\blambda}{}_{\blambda \bi \bj} 
\right) 
\,,
\ee
\be
\widetilde{T}_{\bmu \bi \bj}{}^{\bk} =
e^{\phi/2} \left( 
\hat\partial_{\bi \bj} V_{\bmu}^{\bk} - 2 \tilde\Gamma_{\bmu [\bi\bj]}{}^{\bk} 
+ \frac{2}{3} \delta^{\bk}_{[\bi} \left( \hat \partial_{\bj] \bar{l}} V^{\bar{l}}_{\bmu} 
+ \mathcal{Q}^{\brho}{}_{\bj] \bmu \brho} 
+ \tilde \Gamma_{|\bmu \bar{l} | \bj]}{}^{\bar{l}}
- 2 \nT_{\bj]\bmu} \right)
\right) 
+ V^{\bk}_{\bnu} \widetilde{T}_{\bmu \bi \bj}{}^{\bnu} 
\,,
\ee
\be
\widetilde{T}_{\bmu \bnu \bi}{}^{\brho} = 
e^{\phi/2} \left( 
\mathcal{Q}^{\brho}{}_{\bi \bmu \bnu} 
- \frac{2}{3} \delta^{\brho}_{[\bmu}
\left(
\nT_{|\bi|\bnu]} + \hat\partial_{|\bi\bj|} V_{\bnu]}^{\bj}  
- 2 \mathcal{Q}^{\blambda}{}_{|\bi | \bnu] \blambda }
+ \tilde \Gamma_{\bnu] \bj \bi}{}^{\bj} 
\right) 
\right)
+ 2 V^{\bk}_{[\bmu} \widetilde{T}_{\bnu]\bk\bi}{}^{\brho} \,,
\ee
\be
\begin{split}
\widetilde{T}_{\bmu \bnu \bi}{}^{\bj} & = 
e^{\phi/2} \left( 
2 \tilde\partial_{\bi[\bmu} V_{\bnu]}^{\bj} - \epsilon_{\bmu \bnu \blambda} \Gamma^{\blambda}{}_{\bi}{}^{\bj} 
+ \frac{2}{3} \delta^{\bj}_{\bi} \left(
\frac{1}{2} \epsilon_{\bar{\kappa} \bar{\lambda} [ \bmu} T^{\bar{\kappa} \blambda}{}_{\bnu]}
- \tilde\partial_{\bk [ \bmu} V_{\bnu]}^{\bk} 
\right) 
\right)
\\ & \qquad 
+ 2 V^{\bk}_{[\bmu} \widetilde{T}_{\bnu] \bk \bi}{}^{\bj} 
+ V_{\brho}^{\bj} \widetilde{T}_{\bmu \bnu \bi}{}^{\brho} 
- 2 V_{\brho}^{\bj} V_{[\bmu}^{\bk} \widetilde{T}_{\bnu] \bk \bi}{}^{\brho} \,,
\end{split}
\ee
\be
\begin{split}
\widetilde{T}_{\bmu \bnu \brho}{}^{\blambda} & =
e^{\phi/2} 
\epsilon_{\bmu \bnu \brho} 
\left( 
\frac{2}{3} T^{\blambda \bar{\kappa} }{}_{\bar{\kappa}}
-\frac{1}{3}\epsilon^{\blambda \bar{\kappa}\bar{\sigma}} \tilde\partial_{|\bi|\bar{\kappa}} V_{\bar{\sigma}]}^{\bi} 
\right) \\ & \qquad
-3 V_{[\bmu}^{\bi} \widetilde{T}_{\bnu \brho] \bi }{}^{\blambda}
-3 V^{\bi}_{[\bmu} V^{\bj}_{\bnu} \widetilde{T}_{\brho]\bi\bj}{}^{\blambda} 
 \,,
\end{split}
\ee
\be
\begin{split}
\widetilde{T}_{\bmu \bnu \brho}{}^{\bi} & = e^{\phi/2} \epsilon_{\bmu \bnu \brho} \partial^{\blambda} V_{\blambda}^{\bi}
+ V_{\blambda}^{\bi} \widetilde{T}_{\bmu \bnu \brho}{}^{\blambda}
- 3 V_{[\bmu}^{\bj} \widetilde{T}_{\bnu \brho] \bj}{}^{\bi}
\\ & 
\qquad + 3 V_{\blambda}^{\bi} V_{[\bmu}^{\bj} \widetilde{T}_{\bnu \brho] \bj}{}^{\blambda} 
- 3 V_{[\bmu}^{\bj} V_{\bnu}^{\bk} \widetilde{T}_{\brho] \bj \bk}{}^{\bi} 
+ 3 V_{\blambda}^{\bi} V_{[\bmu}^{\bj} V_{\bnu}^{\bk} \widetilde{T}_{\brho] \bj \bk}{}^{\blambda} \,.
\end{split}
\ee

\bibliographystyle{JHEP}
\bibliography{NewBib}
\end{document}